\documentclass[showpacs,preprintnumbers,aps]{revtex4}
\usepackage{amssymb}

\usepackage{graphicx}
\usepackage{bm}
\usepackage{color}
\usepackage{ulem}

\begin{document}

\title{Spin-orbit coupled Bose-Einstein condensates in a double well}
\author{Roberta Citro}
\affiliation{Dipartimento di Fisica "E. R. Caianiello",
Universit\'{a} degli Studi di Salerno and CNR-SPIN, Unit\'{a} Operativa di Salerno, Via Ponte Don Melillo, 84084 Fisciano (SA),
Italy}
\author{Adele Naddeo}
\affiliation{Dipartimento di Fisica "E. R. Caianiello",
Universit\'{a} degli Studi di Salerno and CNISM, Unit\'{a} di Ricerca di Salerno, Via Ponte Don Melillo, 84084 Fisciano (SA),
Italy}

\date{\today}

\begin{abstract}
We study the quantum dynamics of a spin-orbit (SO) coupled Bose-Einstein
condensate (BEC) in a double-well potential inspired by the experimental protocol recently developed by NIST group. We focus on the
regime where the number of atoms is very large and perform a two-mode
approximation. An analytical solution of the two-site Bose-Hubbard-like
Hamiltonian is found for several limiting cases, which range from a strong
Raman coupling to a strong Josephson coupling, ending with the complete
model in the presence of weak nonlinear interactions.
Depending on the particular limit, different approaches are chosen: a mapping
onto an $SU(2)$ spin problem together with a Holstein-Primakoff
transformation in the first two cases and a rotating wave approximation
(RWA) when dealing with the complete model. The quantum evolution of the
number difference of bosons with equal or different spin between the two wells is
investigated in a wide range of
parameters; finally the corresponding total atomic current and the spin
current are computed. We show a spin Josephson effect
which could be detected in experiments and employed to build up realistic
devices.
\end{abstract}

\maketitle



\section{Introduction}

Spin-orbit coupling relates the velocity of a particle to its spin and is
ubiquitous in condensed matter physics. It plays a key role in a variety of
systems and gives rise to new phenomena ranging from topological insulators
\cite{top1} to spin-Hall effect \cite{top2} and Majorana fermions \cite{top3}%
. In solid-state materials spin-orbit coupling arises because of the motion
of electrons in the intrinsic electric field of the crystal, which is a
characteristics of the material under study. On the other hand ultracold
atomic gases offer an unique platform for engineering synthetic spin-orbit
couplings thanks to the wide tunability of experimental parameters \cite
{cold1}. That is achieved by controlling atom-light interactions and has
recently led to the generation of effective Abelian and non-Abelian gauge
fields \cite{cold2}. In the last years some experimental proposals have been
implemented. In a series of pioneering experiments \cite{nist1} the NIST
group successfully built up synthetic uniform gauge fields, magnetic fields,
electric fields and SO couplings \cite{nist2}. In particular SO coupling
\cite{nist2} with equal Rashba \cite{ra1} and Dresselhaus \cite{ra2}
strengths in a neutral atomic BEC has been engineered by dressing two atomic
spin states with a pair of counterpropagating laser beams. Furthermore laser
coupling has been shown to induce a modification on the dressed spin states
by driving a quantum phase transition from a spatially spin-mixed state to a
phase-separated state. The above scheme has been further generalized to
create nearly isotropic Rashba SO coupling as well as a tunable combination
of Rashba and Dresselhaus SO coupling \cite{campbell}.

Up to now a number of theoretical investigations on SO-coupled BECs has been
performed, concerning phase diagrams in the ground state \cite{th1}\cite
{th1bis}, vortices structures in the presence of external rotation \cite{th2}%
, unconventional collective dipole oscillations \cite{th3}\cite{th3bis},
superfluid to Mott insulator transitions in a lattice \cite{th4}, supersolid
features in the excitation spectrum within the stripe phase \cite{th4bis}
and interesting simulation of relativistic effects such as Zitterbewegung
\cite{th5} and Klein tunneling \cite{th6}. The key consequence of SO
coupling which emerges in all such examples is an enhancement of interaction
effects even for a weak interacting BEC \cite{th7}.

On the other hand, in the last  years great efforts have been devoted
to the exploration of the role of quantum fluctuations and in general of
macroscopic quantum coherence phenomena \cite{leggett1} in BECs, in order to
understand the intriguing interplay between nonlinear interactions and
quantum coherence and the emergent new phenomena which could arise in such a
new environment. The prototypical system one can study is a BEC in a double
well potential, which represents the cold atom analogue of a Josephson
junction \cite{junction1}. Within a mean field approximation a reliable
description can be obtained by means of the Gross-Pitaevskii theory which
gives rise to a variety of phenomena, ranging from Josephson oscillations
\cite{jos1}\cite{milburn}\cite{ananikian1} to macroscopic quantum
self-trapping (MQST) \cite{smerzi1} and ac and dc Josephson like effect \cite
{smerzi2}, all experimentally observed in the last decade \cite{exp1}\cite
{exp2}\cite{exp3}. Vice versa, in a quantum regime and within the tight
binding approximation one gets the Bose-Hubbard dimer Hamiltonian \cite
{dimer1}\cite{gati1}\cite{ferrini1}, whose parameters are the hopping
frequency $J$ \ between the two lattice sites, the onsite interaction
strength $g_{j}$, $j=1,2$ and the total atoms number $N$. Furthermore it can
be mapped onto a $SU(2)$ spin problem which coincides with the
Lipkin-Meshkov-Glick (LMG) model \cite{lipkin1,lipkin2}. More recently, the
theoretical analysis on weakly coupled condensates has been successfully
extended to a binary mixture of BECs in a double well potential \cite{mix0}
\cite{hp3}\cite{mix1}\cite{mix2}\cite{mix3}\cite{mix4}\cite{mix4bis}\cite
{mix5}\cite{noi1}\cite{noi2}, resulting in a richer tunneling dynamics,
which includes two different MQST states with broken symmetry \cite{mix3},
characterized by localization in the two different wells (phase separation)
or coexistence in the same well respectively. Furthermore the coherent
dynamics of a two species BEC in a double well has been analyzed as well
focussing on the case where the two species are two hyperfine states of the
same alkali metal \cite{mix6}.

Till now, the quantum dynamics of SO coupled BECs in a double well is still poorly
investigated. Recent mean-field results \cite{cinesi1} relying on the
experimental setup by NIST group \cite{nist2} point towards an interplay
between external and internal Josephson effects mainly in the absence of
interatomic interactions, as well as towards the existence of a net atomic spin
current in the weak Raman coupling regime. Likewise, in Ref. \cite{MDS1} a classical study of the interplay between interatomic interactions and SO coupling
has been reported as well, together with a careful analysis of the self-trapped dynamics of the total population imbalance between the two bosonic pseudospin species.

In this paper we carry out a comprehensive analytical study of the quantum
behavior of SO coupled BECs by making explicit reference to the experimental
setup by NIST group \cite{nist2} as well. We analyze in detail the weak
interaction limit and start from a two-mode Bose-Hubbard-like Hamiltonian.
We focus on three parameters regimes: weak Raman coupling, strong Raman
coupling, intermediate regime without and with interatomic interactions. The
first two cases are treated by a mapping onto a $SU(2)$ spin problem
together with a Holstein-Primakoff transformation \cite{hp1}\cite{hp2}. As a
result a Hamiltonian of decoupled quantum harmonic oscillators is obtained,
whose stationary states are readily found. Finally, the intermediate regime
case is dealt with starting from the non interacting case where a simple
diagonalization is enough. This study is preliminary but already shows up
some interesting features such as spin currents and spin Josephson-like
effects. Then we switch to the weak interacting regime and obtain a closed
analytical solution via rotating wave approximation \cite{rw1}\cite{rw2}.
For each parameter regime the quantum evolution of the number difference of
bosons of pseudospin up and down between the two wells is investigated in
detail and the total
atomic current and the net spin current can be computed as well.

The paper is organized as follows. In Section 2, by making explicit
reference to the NIST experimental setup, we introduce the model Hamiltonian and focus on the
two mode approximation. In Section 3 we deal with the
two limits of weak and strong Raman couplings. A mapping onto a $SU(2)$ spin
problem together with a Holstein-Primakoff transformation is performed and
the semiclassical limit is taken followed by a decoupling of the bosonic
degrees of freedom. As a result the Hamiltonian can be rephrased in terms of
independent harmonic oscillators, whose stationary states and dynamics is
promptly determined. In Section 4 the intermediate regime is considered both
in the non interacting and in the weakly interacting case and a closed
analytical solution is found. Finally some conclusions are briefly outlined and perspectives of
this work for implementing realistic
devices based on spin Josephson like effects are given.

\section{The model}

In 2011 the NIST group \cite{nist2} succeeded in engineering a SO coupling
with equal Rashba and Dresselhaus strengths in a neutral atomic $^{87}Rb$
BEC by dressing two atomic spin states with a pair of lasers. The key step
in the experimental technique is to select out two internal spin states
within the $F=1$ ground electronic manifold, pseudospin up $\left| \uparrow
\right\rangle =\left| F=1,m_{F}=0\right\rangle $ and pseudospin down $\left|
\downarrow \right\rangle =\left| F=1,m_{F}=-1\right\rangle $, and then
couple them with strength $\Omega $ via a pair of $\lambda _{L}=804.1nm$
Raman lasers, intersecting at an angle $\theta =90^{%
{{}^\circ}%
}$ and detuned by $\delta $ from Raman resonance. Assuming $\hbar k_{L}=%
\sqrt{2}\pi \frac{\hbar }{\lambda _{L}}$ and $E_{L}=\frac{\hbar ^{2}k_{L}^{2}%
}{2m}$ as momentum and energy units, $k_{L}$ being the wave number of the
Raman laser, the SO coupling is described in terms of a single-particle
Hamiltonian:
\begin{equation}
\widehat{H}=\frac{\hbar ^{2}\mathbf{k}^{2}}{2m}I_{2\times 2}+\frac{\delta }{2%
}\sigma _{z}+\frac{\Omega }{2}\sigma _{x}\cos \left( 2k_{L}x\right) -\frac{%
\Omega }{2}\sigma _{y}\sin \left( 2k_{L}x\right) ,  \label{eq1}
\end{equation}
where $\mathbf{k}$ is the atomic momentum in the $x-y$ plane, $m$ is the
atomic mass, $I_{2\times 2}$ is the identity matrix and $\sigma _{x}$, $%
\sigma _{y}$, $\sigma _{z}$ are the $2\times 2$ Pauli matrices. Since SO
coupling acts only in one spatial dimension, in the following we neglect the
motion of atoms along $y$ and $z$ axis and consider Eq. (\ref{eq1})
restricted to $x$ axis.

After the transformation $U\equiv \left(
\begin{array}{cc}
e^{-ik_{L}x} & 0 \\
0 & e^{ik_{L}x}
\end{array}
\right) $ within the space $\left| \uparrow \right\rangle $, $\left|
\downarrow \right\rangle $, dressed pseudospins $\left| \uparrow
\right\rangle _{d}=e^{-ik_{L}x}\left| \uparrow \right\rangle $, $\left|
\downarrow \right\rangle _{d}=e^{ik_{L}x}\left| \downarrow \right\rangle $
are introduced and the one-dimensional version of Hamiltonian (\ref{eq1})
takes the form:
\begin{equation}
\widehat{H}_{d}=\frac{\hbar ^{2}\widehat{k}_{x}^{2}}{2m}I_{2\times
2}+2\alpha \widehat{k}_{x}\sigma _{z}+\frac{\Omega }{2}\sigma _{x}+\frac{%
\delta }{2}\sigma _{z},  \label{eq2}
\end{equation}
where $\alpha =\frac{E_{L}}{k_{L}}$. Let us now notice that, for $\delta =0$%
, Eq. (\ref{eq2}) gives rise to the following dispersion relation
\begin{equation}
E_{\pm }\left( k_{x}\right) =\frac{\hbar ^{2}\widehat{k}_{x}^{2}}{2m}\pm
\sqrt{4\alpha ^{2}k_{x}^{2}+\frac{\Omega ^{2}}{4}},  \label{eq3}
\end{equation}
which shows two branches. The lowest one, for $\Omega <4E_{L}$, exhibits a
double well structure with two minima corresponding to the condensation of
dressed pseudospin up and down states, while Raman coupling and a small
detuning $\delta $ modulate the atomic population in the above two states.
The complete experimental control and wide tunability of $\delta $, $\Omega $
and $k_{L}$ parameters in Eq. (\ref{eq2}) allows one to select out the
region $\Omega <4E_{L}$ within the parameters space, thus in the following
we will work in the dressed pseudospin basis $\left| \uparrow \right\rangle
_{d}$, $\left| \downarrow \right\rangle _{d}$ and focus on such a regime.

Now let us switch interatomic interactions and put the SO coupled BEC in a
spin independent double well trapping potential $V\left( x\right) $. The
second quantized version of the full Hamiltonian takes the form:
\begin{equation}
\mathcal{H}=\mathcal{H}_{0}+\mathcal{H}_{int},  \label{eq4}
\end{equation}
where
\begin{equation}
\mathcal{H}_{0}=\int dx\widehat{\Psi }^{\dagger }\left( x\right) \left[
\widehat{H}_{d}+V\left( x\right) \right] \widehat{\Psi }\left( x\right)
=\int dx\left(
\begin{array}{cc}
\widehat{\Psi }_{\uparrow }^{\dagger }\left( x\right) & \widehat{\Psi }%
_{\downarrow }^{\dagger }\left( x\right)
\end{array}
\right) \left(
\begin{array}{cc}
H_{\uparrow }+V\left( x\right) & \frac{\Omega }{2} \\
\frac{\Omega }{2} & H_{\downarrow }+V\left( x\right)
\end{array}
\right) \left(
\begin{array}{c}
\widehat{\Psi }_{\uparrow }\left( x\right) \\
\widehat{\Psi }_{\downarrow }\left( x\right)
\end{array}
\right) ,  \label{eq5}
\end{equation}
and
\begin{equation}
\mathcal{H}_{int}=\frac{g_{\uparrow \uparrow }}{2}\int dx\widehat{\Psi }%
_{\uparrow }^{\dagger }\left( x\right) \widehat{\Psi }_{\uparrow }^{\dagger
}\left( x\right) \widehat{\Psi }_{\uparrow }\left( x\right) \widehat{\Psi }%
_{\uparrow }\left( x\right) +\frac{g_{\downarrow \downarrow }}{2}\int dx%
\widehat{\Psi }_{\downarrow }^{\dagger }\left( x\right) \widehat{\Psi }%
_{\downarrow }^{\dagger }\left( x\right) \widehat{\Psi }_{\downarrow }\left(
x\right) \widehat{\Psi }_{\downarrow }\left( x\right) +g_{\uparrow
\downarrow }\int dx\widehat{\Psi }_{\uparrow }^{\dagger }\left( x\right)
\widehat{\Psi }_{\downarrow }^{\dagger }\left( x\right) \widehat{\Psi }%
_{\uparrow }\left( x\right) \widehat{\Psi }_{\downarrow }\left( x\right) .
\label{eq6}
\end{equation}
Here $H_{\uparrow }=\frac{\hbar ^{2}}{2m}\left( \widehat{k}_{x}^{2}+2k_{L}%
\widehat{k}_{x}\right) +\frac{\delta }{2}$, $H_{\downarrow }=\frac{\hbar ^{2}%
}{2m}\left( \widehat{k}_{x}^{2}-2k_{L}\widehat{k}_{x}\right) -\frac{\delta }{%
2}$ and $g_{\sigma \sigma ^{\prime }}=\frac{2\hbar ^{2}a_{\sigma \sigma
^{\prime }}}{ml_{\perp }^{2}}$ with $\sigma ,\sigma ^{\prime }=\uparrow
,\downarrow $ is the interaction strength, $a_{\sigma ,\sigma ^{\prime }}$
being the $s$-wave scattering length between pseudospin $\sigma $ and $%
\sigma ^{\prime }$ and $l_{\perp }$ the oscillator length due to a harmonic
vertical confinement; furthermore $\widehat{\Psi }_{\sigma }^{\dagger
}\left( x\right) ,$ $\widehat{\Psi }_{\sigma }\left( x\right) $, $\sigma
=\uparrow ,\downarrow $ are the bosonic field operators, which satisfy the
commutation rules:
\begin{eqnarray}
\left[ \widehat{\Psi }_{\sigma }\left( x\right) ,\widehat{\Psi }_{\sigma
^{\prime }}\left( x^{\prime }\right) \right] &=&\left[ \widehat{\Psi }%
_{\sigma }^{\dagger }\left( x\right) ,\widehat{\Psi }_{\sigma ^{\prime
}}^{\dagger }\left( x^{\prime }\right) \right] =0,  \label{eq7} \\
\left[ \widehat{\Psi }_{\sigma }\left( x\right) ,\widehat{\Psi }_{\sigma
^{\prime }}^{\dagger }\left( x^{\prime }\right) \right] &=&\delta _{\sigma
\sigma ^{\prime }}\delta \left( x-x^{\prime }\right) ,\text{ \ \ \ \ }\sigma
,\sigma ^{\prime }=\uparrow ,\downarrow ,  \label{eq8}
\end{eqnarray}
and the normalization conditions:
\begin{equation}
\int dx\left| \widehat{\Psi }_{\sigma }\left( x\right) \right|
^{2}=N_{\sigma };\text{ \ \ \ }\sigma =\uparrow ,\downarrow ,  \label{eq9}
\end{equation}
$N_{\sigma }$, $\sigma =\uparrow ,\downarrow $ being the number of atoms
with pseudospin $\sigma $ and $\sigma ^{\prime }$ respectively. The total
number of atoms in the system is $N=N_{\uparrow }+N_{\downarrow }$.

A weak link between the two wells produces a small energy splitting between
the mean-field ground state and the first excited state of the double well
potential and that allows to reduce the dimension of the Hilbert space of
the initial many-body problem. Indeed for low energy excitations, low
temperatures and a small effective Zeeman splitting it is possible to
consider only such two states and neglect the contribution from the higher
ones, in this way performing a two-mode approximation \cite{milburn}\cite
{smerzi1}\cite{ananikian1}. As a consequence, the field operator can be
expressed as:
\begin{equation}
\begin{array}{cc}
\widehat{\Psi }_{\sigma }\left( x\right) \simeq a_{L\sigma }\psi _{L\sigma
}\left( x\right) +a_{R\sigma }\psi _{R\sigma }\left( x\right) , & \sigma
=\uparrow ,\downarrow
\end{array}
\label{eq10}
\end{equation}
where $\psi _{j\sigma }\left( x\right) $, $j=L,R$, is the ground state wave
function in the $j$ well with pseudospin $\sigma $ and $a_{j\sigma }$ is the
corresponding annihilation operator, which obeys to the bosonic commutation
relation $\left[ a_{j\sigma },a_{k\sigma ^{\prime }}^{\dagger }\right]
=\delta _{jk}\delta _{\sigma \sigma ^{\prime }}$.

By putting Eq. (\ref{eq10}) in Eqs. (\ref{eq4})-(\ref{eq6}) and neglecting
interwell atomic interactions as well as two-particle processes we turn the
total Hamiltonian $\mathcal{H}=\mathcal{H}_{0}+\mathcal{H}_{int}$ into the
following Bose-Hubbard like form:
\begin{eqnarray}
\mathcal{H} &=&\sum_{j=L,R}\sum_{\sigma =\uparrow ,\downarrow }\varepsilon
_{j\sigma }a_{j\sigma }^{\dagger }a_{j\sigma }+\sum_{\sigma ,\sigma ^{\prime
}=\uparrow ,\downarrow }\left( J_{\sigma \sigma ^{\prime }}a_{L\sigma
}^{\dagger }a_{R\sigma ^{\prime }}+J_{\sigma \sigma ^{\prime }}^{\ast
}a_{R\sigma ^{\prime }}^{\dagger }a_{L\sigma }\right)  \nonumber \\
&&+\frac{1}{2}\sum_{j=L,R}\left( \Omega _{j}a_{j\uparrow }^{\dagger
}a_{j\downarrow }+\Omega _{j}^{\ast }a_{j\downarrow }^{\dagger }a_{j\uparrow
}\right) +\frac{\delta }{2}\sum_{j=L,R}\left( a_{j\uparrow }^{\dagger
}a_{j\uparrow }-a_{j\downarrow }^{\dagger }a_{j\downarrow }\right)
\label{eq11} \\
&&+\frac{1}{2}\sum_{j=L,R}\left( g_{\uparrow \uparrow }^{\left( j\right)
}a_{j\uparrow }^{\dagger }a_{j\uparrow }^{\dagger }a_{j\uparrow
}a_{j\uparrow }+g_{\downarrow \downarrow }^{\left( j\right) }a_{j\downarrow
}^{\dagger }a_{j\downarrow }^{\dagger }a_{j\downarrow }a_{j\downarrow
}+2g_{\uparrow \downarrow }^{\left( j\right) }a_{j\uparrow }^{\dagger
}a_{j\downarrow }^{\dagger }a_{j\uparrow }a_{j\downarrow }\right) .
\nonumber
\end{eqnarray}
Here $\varepsilon _{j\uparrow }=\int dx\psi _{j\uparrow }^{\ast }\left(
x\right) \left[ \frac{\hbar ^{2}}{2m}\left( \widehat{k}_{x}^{2}+2k_{L}%
\widehat{k}_{x}\right) +V\left( x\right) \right] \psi _{j\uparrow }\left(
x\right) $ and $\varepsilon _{j\downarrow }=\int dx\psi _{j\uparrow }^{\ast
}\left( x\right) \left[ \frac{\hbar ^{2}}{2m}\left( \widehat{k}%
_{x}^{2}-2k_{L}\widehat{k}_{x}\right) +V\left( x\right) \right] \psi
_{j\downarrow }\left( x\right) $ are the single-particle ground state
energies in the well $j$, $J_{\sigma \sigma }=\int dx\psi _{L\sigma }^{\ast
}\left( x\right) \left[ H_{\sigma }+V\left( x\right) \right] \psi _{R\sigma
}\left( x\right) $ are the Josephson tunneling terms between left and right
well, $J_{\sigma \overline{\sigma }}=\int dx\psi _{L\sigma }^{\ast }\left(
x\right) \frac{\Omega }{2}\psi _{R\overline{\sigma }}\left( x\right) $ with
different spins $\sigma $ and $\overline{\sigma }$ are the interwell
spin-flip tunneling terms induced by the Raman coupling, $\Omega _{j}=\Omega
\int dx\psi _{j\uparrow }^{\ast }\left( x\right) \psi _{j\downarrow }\left(
x\right) $ are the Raman coupling terms in each well and, finally, $%
g_{\sigma \sigma ^{\prime }}^{\left( j\right) }=g_{\sigma \sigma ^{\prime
}}\int dx\left| \psi _{j\sigma }\left( x\right) \right| ^{2}\left| \psi
_{j\sigma ^{\prime }}\left( x\right) \right| ^{2}$ is the effective
interaction strength.

The complete control of the experimental environment makes possible a fine
tuning of the parameters. In this way a symmetric double-well potential can
be realized, which allows one to set $\varepsilon _{L\uparrow }=\varepsilon
_{R\uparrow }=\varepsilon _{L\downarrow }=\varepsilon _{R\downarrow }\equiv
\varepsilon $, $\Omega _{L}=\Omega _{R}^{\ast }\simeq \Omega _{L}^{\ast
}=\Omega _{R}\equiv \overline{\Omega }$ and $g_{\sigma \sigma ^{\prime
}}^{\left( L\right) }=g_{\sigma \sigma ^{\prime }}^{\left( R\right) }\equiv
\overline{g}_{\sigma \sigma ^{\prime }}$. Also the spin-flip tunneling
amplitude $J_{\sigma \overline{\sigma }}$ can be dropped under realistic
experimental conditions \cite{nist2}. Further simplifying assumptions amount
to neglect the constant energy shifts $\varepsilon \left( N_{L\uparrow
}+N_{L\downarrow }+N_{R\uparrow }+N_{R\downarrow }\right) $, so that the
Hamiltonian (\ref{eq11}) becomes:
\begin{eqnarray}
\mathcal{H} &=&J_{\uparrow \uparrow }\left( a_{L\uparrow }^{\dagger
}a_{R\uparrow }+a_{R\uparrow }^{\dagger }a_{L\uparrow }\right)
+J_{\downarrow \downarrow }\left( a_{L\downarrow }^{\dagger }a_{R\downarrow
}+a_{R\downarrow }^{\dagger }a_{L\downarrow }\right) +\frac{\overline{\Omega
}}{2}\left( a_{L\uparrow }^{\dagger }a_{L\downarrow }+a_{L\downarrow
}^{\dagger }a_{L\uparrow }+a_{R\uparrow }^{\dagger }a_{R\downarrow
}+a_{R\downarrow }^{\dagger }a_{R\uparrow }\right)  \nonumber \\
&&+\frac{\delta }{2}\left( a_{L\uparrow }^{\dagger }a_{L\uparrow
}+a_{R\uparrow }^{\dagger }a_{R\uparrow }-a_{L\downarrow }^{\dagger
}a_{L\downarrow }-a_{R\downarrow }^{\dagger }a_{R\downarrow }\right) +\frac{1%
}{2}\overline{g}_{\uparrow \uparrow }\left( a_{L\uparrow }^{\dagger
}a_{L\uparrow }^{\dagger }a_{L\uparrow }a_{L\uparrow }+a_{R\uparrow
}^{\dagger }a_{R\uparrow }^{\dagger }a_{R\uparrow }a_{R\uparrow }\right)
\label{eq12} \\
&&+\frac{1}{2}\overline{g}_{\downarrow \downarrow }\left( a_{L\downarrow
}^{\dagger }a_{L\downarrow }^{\dagger }a_{L\downarrow }a_{L\downarrow
}+a_{R\downarrow }^{\dagger }a_{R\downarrow }^{\dagger }a_{R\downarrow
}a_{R\downarrow }\right) +\overline{g}_{\uparrow \downarrow }\left(
a_{L\uparrow }^{\dagger }a_{L\downarrow }^{\dagger }a_{L\uparrow
}a_{L\downarrow }+a_{R\uparrow }^{\dagger }a_{R\downarrow }^{\dagger
}a_{R\uparrow }a_{R\downarrow }\right) .  \nonumber
\end{eqnarray}

Finally, let us give some orders of magnitude estimations for the parameters
appearing in Eq. (\ref{eq12}) by making explicit reference to the
experimental setup of Ref.\onlinecite{nist2}. Indeed for the Raman
lasers we assume a wavelength $\lambda _{L}=804.1nm$ and a recoil frequency $%
\frac{E_{L}}{\hbar }=22.5kHz$ and choose $\Omega $ (that is $\overline{%
\Omega }$) in such a way to fulfil the condition $\Omega <4E_{L}$. The
energy scale of the Zeeman field $\delta $ generally satisfies the condition
$\delta <<E_{L}$ and eventually gets the limiting value $0.01E_{L}$ while
the tunneling terms may be chosen as $J_{\uparrow \uparrow },J_{\downarrow
\downarrow }\approx -0.1E_{L}$ \cite{cinesi1}. Finally, for a $^{87}Rb$ BEC
the trapping frequency of each well could be $\omega \sim 0.1\frac{E_{L}}{%
\hbar }$ \cite{exp2}. In the following we will always consider a weak
nonlinear interaction, which could be easily obtained by means of Feshbach
resonances technique, so that the condition $\overline{g}_{\uparrow \uparrow
},\overline{g}_{\downarrow \downarrow },\overline{g}_{\uparrow \downarrow
}<<\hbar \omega $ holds and gives $\overline{g}_{\uparrow \uparrow },%
\overline{g}_{\downarrow \downarrow },\overline{g}_{\uparrow \downarrow
}<<0.1E_{L}$.

In the following Sections we study this Hamiltonian in some limiting cases,
amenable to analytical solutions. In the most general case, when parameters
vary in a wide and arbitrary range, an analytical solution in closed form cannot
be found and one has to resort to numerical calculations. The numerical
solution of this problem will be the subject of a future publication \cite
{noiF}.

\section{Quantum dynamics within weak and strong Raman coupling regimes}

In this Section we start the analysis of the model Hamiltonian (\ref{eq12})
by focusing on two simple limiting cases, the weak and strong Raman
coupling regime, respectively. A closed form analytical solution is obtained
in both limits, by adopting the same procedure as in Ref. \cite{noi2}. We
perform a mapping onto a $SU(2)$ spin problem together with a
Holstein-Primakoff transformation; in this way the semiclassical limit is
taken followed by a decoupling of the bosonic degrees of freedom. As a
result the Hamiltonian will be rephrased in terms of independent harmonic
oscillators, whose stationary states and dynamics is easily established.

\subsection{Weak Raman coupling regime}

This limiting case corresponds to the physical situation $\frac{\overline{%
\Omega }}{E_{L}}<<1$, $\frac{\overline{\Omega }}{|J_{\uparrow \uparrow }|}<<1$
and $\frac{\overline{\Omega }}{|J_{\downarrow \downarrow }|}<<1$, so that the
relevant dynamics is governed by an external Josephson-like effect.
Spin-flip processes can be safely neglected while we get two Josephson
tunneling processes, one for each pseudospin. As a consequence the total
number of particles is conserved for pseudospin up and down, $N_{\uparrow
}=N_{L\uparrow }+N_{R\uparrow }$ and $N_{\downarrow }=N_{L\downarrow
}+N_{R\downarrow }$, respectively. Furthermore in the following the simplest
assumption $N_{\uparrow }=N_{\downarrow }=\frac{N}{2}$ will be taken.

In this limit Hamiltonian (\ref{eq12}) reduces to:
\begin{eqnarray}
\mathcal{H} &=&J_{\uparrow \uparrow }\left( a_{L\uparrow }^{\dagger
}a_{R\uparrow }+a_{R\uparrow }^{\dagger }a_{L\uparrow }\right)
+J_{\downarrow \downarrow }\left( a_{L\downarrow }^{\dagger }a_{R\downarrow
}+a_{R\downarrow }^{\dagger }a_{L\downarrow }\right) +\frac{\delta }{2}%
\left( a_{L\uparrow }^{\dagger }a_{L\uparrow }+a_{R\uparrow }^{\dagger
}a_{R\uparrow }-a_{L\downarrow }^{\dagger }a_{L\downarrow }-a_{R\downarrow
}^{\dagger }a_{R\downarrow }\right)  \nonumber \\
&&+\frac{1}{2}\overline{g}_{\uparrow \uparrow }\left( a_{L\uparrow
}^{\dagger }a_{L\uparrow }^{\dagger }a_{L\uparrow }a_{L\uparrow
}+a_{R\uparrow }^{\dagger }a_{R\uparrow }^{\dagger }a_{R\uparrow
}a_{R\uparrow }\right) +\frac{1}{2}\overline{g}_{\downarrow \downarrow
}\left( a_{L\downarrow }^{\dagger }a_{L\downarrow }^{\dagger }a_{L\downarrow
}a_{L\downarrow }+a_{R\downarrow }^{\dagger }a_{R\downarrow }^{\dagger
}a_{R\downarrow }a_{R\downarrow }\right)  \label{eq13} \\
&&+\overline{g}_{\uparrow \downarrow }\left( a_{L\uparrow }^{\dagger
}a_{L\downarrow }^{\dagger }a_{L\uparrow }a_{L\downarrow }+a_{R\uparrow
}^{\dagger }a_{R\downarrow }^{\dagger }a_{R\uparrow }a_{R\downarrow }\right)
.  \nonumber
\end{eqnarray}
When introducing the angular momentum operators for pseudospin $\uparrow $
and $\downarrow $:
\begin{equation}
\begin{array}{ccc}
J_{x}^{\uparrow }=\frac{1}{2}\left( a_{R\uparrow }^{\dagger }a_{L\uparrow
}+a_{L\uparrow }^{\dagger }a_{R\uparrow }\right) , & J_{y}^{\uparrow }=\frac{%
i}{2}\left( a_{R\uparrow }^{\dagger }a_{L\uparrow }-a_{L\uparrow }^{\dagger
}a_{R\uparrow }\right) , & J_{z}^{\uparrow }=\frac{1}{2}\left( a_{R\uparrow
}^{\dagger }a_{R\uparrow }-a_{L\uparrow }^{\dagger }a_{L\uparrow }\right) ,
\\
J_{x}^{\downarrow }=\frac{1}{2}\left( a_{R\downarrow }^{\dagger
}a_{L\downarrow }+a_{L\downarrow }^{\dagger }a_{R\downarrow }\right) , &
J_{y}^{\downarrow }=\frac{i}{2}\left( a_{R\downarrow }^{\dagger
}a_{L\downarrow }-a_{L\downarrow }^{\dagger }a_{R\downarrow }\right) &
J_{z}^{\downarrow }=\frac{1}{2}\left( a_{R\downarrow }^{\dagger
}a_{R\downarrow }-a_{L\downarrow }^{\dagger }a_{L\downarrow }\right) ,
\end{array}
\label{eq14}
\end{equation}
where the operators $J_{i}^{\uparrow }$, $J_{i}^{\downarrow }$, $i=x,y,z$,
obey to the usual angular momentum algebra together with the relation:
\begin{equation}
\begin{array}{cc}
\left( J^{\uparrow }\right) ^{2}=\frac{N_{\uparrow }}{2}\left( \frac{%
N_{\uparrow }}{2}+1\right) , & \left( J^{\downarrow }\right) ^{2}=\frac{%
N_{\downarrow }}{2}\left( \frac{N_{\downarrow }}{2}+1\right) ,
\end{array}
\label{eq15}
\end{equation}
Hamiltonian (\ref{eq13}) reduces to a sum of two Lipkin-Meshkov-Glick (LMG)
models \cite{lipkin1,lipkin2}:
\begin{equation}
\mathcal{H}=\frac{\delta }{4}\left( N_{L\uparrow }-N_{L\downarrow
}+N_{R\uparrow }-N_{R\downarrow }\right) +\overline{g}_{\uparrow \uparrow
}\left( J_{z}^{\uparrow }\right) ^{2}+J_{\uparrow \uparrow }J_{x}^{\uparrow
}+\overline{g}_{\downarrow \downarrow }\left( J_{z}^{\downarrow }\right)
^{2}+J_{\downarrow \downarrow }J_{x}^{\downarrow }+2\overline{g}_{\uparrow
\downarrow }J_{z}^{\uparrow }J_{z}^{\downarrow },  \label{eq16}
\end{equation}
which are coupled for $\overline{g}_{\uparrow \downarrow }\neq 0$.

To proceed further, let us focus on the regime with large number of atoms $%
N_{\uparrow },N_{\downarrow }\gg 1$ and weak scattering strengths $%
J_{\uparrow \uparrow },J_{\downarrow \downarrow }\gg \overline{g}_{\uparrow
\uparrow },\overline{g}_{\downarrow \downarrow },\overline{g}_{\uparrow
\downarrow }$ and make the rotation:
\begin{equation}
\begin{array}{cc}
\begin{array}{c}
J_{z}^{\uparrow }\rightarrow -J_{x}^{\uparrow } \\
J_{x}^{i}\rightarrow J_{z}^{i}
\end{array}
, & i=\uparrow ,\downarrow ,
\end{array}
\label{eq17}
\end{equation}
followed by a linearized Holstein-Primakoff transformation \cite{hp1,hp2}
\begin{equation}
\begin{array}{c}
J_{z}^{\uparrow }=J^{\uparrow }-a_{\uparrow }^{\dagger }a_{\uparrow } \\
J_{+}^{\uparrow }=\sqrt{2J^{\uparrow }}a_{\uparrow } \\
J_{-}^{\uparrow }=a_{\uparrow }^{\dagger }\sqrt{2J^{\uparrow }}
\end{array}
,
\begin{array}{c}
J_{z}^{\downarrow }=J^{\downarrow }-a_{\downarrow }^{\dagger }a_{\downarrow }
\\
J_{+}^{\downarrow }=\sqrt{2J^{\downarrow }}a_{\downarrow } \\
J_{-}^{\downarrow }=a_{\downarrow }^{\dagger }\sqrt{2J^{\downarrow }}
\end{array}
,  \label{eq18}
\end{equation}
where $J_{\pm }^{i}=J_{x}^{i}\pm iJ_{y}^{i}$, $J^{i}=N_{i}/2$,$\;i=\uparrow
,\downarrow $, thus leading to the effective Hamiltonian:
\begin{eqnarray}
\mathcal{H} &=&\frac{\delta }{2}\left( J^{\uparrow }-J^{\downarrow }\right)
+2\overline{g}_{\uparrow \uparrow }J^{\uparrow }\left( \frac{a_{\uparrow
}+a_{\uparrow }^{\dagger }}{2}\right) \left( \frac{a_{\uparrow }+a_{\uparrow
}^{\dagger }}{2}\right) +2\overline{g}_{\downarrow \downarrow }J^{\downarrow
}\left( \frac{a_{\downarrow }+a_{\downarrow }^{\dagger }}{2}\right) \left(
\frac{a_{\downarrow }+a_{\downarrow }^{\dagger }}{2}\right) +  \nonumber \\
&&4\overline{g}_{\uparrow \downarrow }\sqrt{J^{\uparrow }J^{\downarrow }}%
\left( \frac{a_{\uparrow }+a_{\uparrow }^{\dagger }}{2}\right) \left( \frac{%
a_{\downarrow }+a_{\downarrow }^{\dagger }}{2}\right) +J_{\uparrow \uparrow
}J^{\uparrow }+J_{\downarrow \downarrow }J^{\downarrow }-J_{\uparrow
\uparrow }a_{\uparrow }^{\dagger }a_{\uparrow }-J_{\downarrow \downarrow
}a_{\downarrow }^{\dagger }a_{\downarrow }.  \label{eq19}
\end{eqnarray}
In this way a mapping from angular momentum operators into bosonic ones has
been constructed. Let us notice that for $\overline{g}_{\uparrow \downarrow
}\neq 0$ the physical Hilbert space breaks into two different sectors
depending on the parity of $a_{\uparrow }^{\dagger }a_{\uparrow
}+a_{\downarrow }^{\dagger }a_{\downarrow }$ and is restricted to $0\leq
a_{\uparrow }^{\dagger }a_{\uparrow }\leq N_{\uparrow }$ and $0\leq
a_{\downarrow }^{\dagger }a_{\downarrow }\leq N_{\downarrow }$.

In order to decouple the degrees of freedom of each bosonic peseudospin
species let us introduce the following harmonic oscillator coordinates and
momenta, $q_{i}$, $p_{i}$, $i=\uparrow ,\downarrow $:
\begin{equation}
\begin{array}{cc}
q_{\uparrow }=\frac{1}{\sqrt{2}}\left( a_{\uparrow }+a_{\uparrow }^{\dagger
}\right) , & q_{\downarrow }=\frac{1}{\sqrt{2}}\left( a_{\downarrow
}+a_{\downarrow }^{\dagger }\right) \\
p_{\uparrow }=\frac{-i}{\sqrt{2}}\left( a_{\uparrow }-a_{\uparrow }^{\dagger
}\right) & p_{\downarrow }=\frac{-i}{\sqrt{2}}\left( a_{\downarrow
}-a_{\downarrow }^{\dagger }\right)
\end{array}
,  \label{eq20}
\end{equation}
which satisfy the usual commutation rules $\left[ q_{i},p_{j}\right]
=i\delta _{ij}$, $i,j=\uparrow ,\downarrow $. Then, by defining:
\begin{equation}
\begin{array}{cc}
Q_{\uparrow }=\frac{q_{\uparrow }}{\sqrt{-J_{\uparrow \uparrow }}}, &
Q_{\downarrow }=\frac{q_{\downarrow }}{\sqrt{-J_{\downarrow \downarrow }}},
\\
P_{\uparrow }=\sqrt{-J_{\uparrow \uparrow }}p_{\uparrow }, & P_{\downarrow }=%
\sqrt{-J_{\downarrow \downarrow }}p_{\downarrow },
\end{array}
\label{eq21}
\end{equation}
(where $\left[ Q_{i},P_{j}\right] =i\delta _{ij}$, $i,j=\uparrow ,\downarrow
$) and, by dropping constant terms $C.st=\left( J_{\uparrow \uparrow }+%
\frac{\delta }{2}\right) J^{\uparrow }+\left( J_{\downarrow \downarrow }-%
\frac{\delta }{2}\right) J^{\downarrow }-\frac{1}{2}\left( J_{\uparrow
\uparrow }+J_{\downarrow \downarrow }\right) $, Eq. (\ref{eq19}) can be
written in a matrix form as:
\begin{equation}
\mathcal{H}\simeq \frac{1}{2}\left[ \hat{Q}^{T}\widehat{\omega }^{2}\hat{Q}+%
\hat{P}^{T}\hat{P}\right] ,  \label{eq22}
\end{equation}
where
\begin{equation}
\widehat{\omega }^{2}=\left(
\begin{array}{cc}
\omega _{\uparrow }^{2} & \omega _{\uparrow \downarrow } \\
\omega _{\uparrow \downarrow } & \omega _{\downarrow }^{2}
\end{array}
\right)  \label{eq23}
\end{equation}
and $\hat{Q}^{T}=(Q_{\uparrow },Q_{\downarrow })$, $\hat{P}^{T}=(P_{\uparrow
},P_{\downarrow })$ (the symbol $\cdot ^{T}$ stands for the transpose);
furthermore the frequencies are $\omega _{\uparrow }^{2}=J_{\uparrow
\uparrow }^{2}-2\overline{g}_{\uparrow \uparrow }J^{\uparrow }J_{\uparrow
\uparrow }$, $\omega _{\uparrow }^{2}=J_{\downarrow \downarrow }^{2}-2%
\overline{g}_{\downarrow \downarrow }J^{\downarrow }J_{\downarrow \downarrow
}$ and $\omega _{\uparrow \downarrow }=2\overline{g}_{\uparrow \downarrow }%
\sqrt{J^{\uparrow }J^{\downarrow }J_{\uparrow \uparrow }J_{\downarrow
\downarrow }}$.

A straightforward diagonalization gives the Hamiltonian:
\begin{equation}
\mathcal{H}\simeq \frac{1}{2}\left[ \omega
_{1}^{2}Q_{1}^{2}+P_{1}^{2}+\omega _{2}^{2}Q_{2}^{2}+P_{2}^{2}\right] ,
\label{eq24}
\end{equation}
where, defining $\Delta _{\uparrow \downarrow }=\sqrt{\left( \omega
_{\uparrow }^{2}-\omega _{\downarrow }^{2}\right) ^{2}+4\omega _{\uparrow
\downarrow }^{2}}$,
\begin{equation}
\begin{array}{cc}
\omega _{1}^{2}=\frac{\omega _{\uparrow }^{2}+\omega _{\downarrow
}^{2}-\Delta _{\uparrow \downarrow }}{2}, & \omega _{2}^{2}=\frac{\omega
_{\uparrow }^{2}+\omega _{\downarrow }^{2}+\Delta _{\uparrow \downarrow }}{2}
\end{array}
,  \label{eq25}
\end{equation}
\begin{equation}
\begin{array}{cc}
Q_{1}=\frac{\left\{ 2\omega _{\uparrow \downarrow }Q_{\downarrow }-\left[
\left( \omega _{\downarrow }^{2}-\omega _{\uparrow }^{2}\right) +\Delta
_{\uparrow \downarrow }\right] Q_{\uparrow }\right\} }{\sqrt{4\omega
_{\uparrow \downarrow }^{2}+\left[ \left( \omega _{\downarrow }^{2}-\omega
_{\uparrow }^{2}\right) +\Delta _{\uparrow \downarrow }\right] ^{2}}}, &
Q_{2}=\frac{\left\{ 2\omega _{\uparrow \downarrow }Q_{\downarrow }-\left[
\left( \omega _{\downarrow }^{2}-\omega _{\uparrow }^{2}\right) -\Delta
_{\uparrow \downarrow }\right] Q_{\uparrow }\right\} }{\sqrt{4\omega
_{\uparrow \downarrow }^{2}+\left[ \left( \omega _{\downarrow }^{2}-\omega
_{\uparrow }^{2}\right) -\Delta _{\uparrow \downarrow }\right] ^{2}}},
\end{array}
\label{eq26}
\end{equation}
\begin{equation}
\begin{array}{cc}
P_{1}=\frac{\left\{ 2\omega _{\uparrow \downarrow }P_{\downarrow }-\left[
\left( \omega _{\downarrow }^{2}-\omega _{\uparrow }^{2}\right) +\Delta
_{\uparrow \downarrow }\right] P_{\uparrow }\right\} }{\sqrt{4\omega
_{\uparrow \downarrow }^{2}+\left[ \left( \omega _{\downarrow }^{2}-\omega
_{\uparrow }^{2}\right) +\Delta _{\uparrow \downarrow }\right] ^{2}}}, &
P_{2}=\frac{\left\{ 2\omega _{\uparrow \downarrow }P_{\downarrow }-\left[
\left( \omega _{\downarrow }^{2}-\omega _{\uparrow }^{2}\right) -\Delta
_{\uparrow \downarrow }\right] P_{\uparrow }\right\} }{\sqrt{4\omega
_{\uparrow \downarrow }^{2}+\left[ \left( \omega _{\downarrow }^{2}-\omega
_{\uparrow }^{2}\right) -\Delta _{\uparrow \downarrow }\right] ^{2}}}.
\end{array}
\label{eq27}
\end{equation}
The symmetric case $\overline{g}_{\uparrow \uparrow }=\overline{g}%
_{\downarrow \downarrow }=\overline{g}$, $J_{\uparrow \uparrow
}=J_{\downarrow \downarrow }=\overline{J}$, $N_{\uparrow }=N_{\downarrow }=%
\frac{N}{2}$ is the simplest, in that we have $\omega _{\uparrow
}^{2}=\omega _{\downarrow }^{2}=\omega ^{2}$ where $\omega ^{2}=\overline{J}%
^{2}-\overline{g}\frac{N}{2}\overline{J}$, and $\omega _{\uparrow \downarrow
}=\overline{g}_{\uparrow \downarrow }\frac{N}{2}\overline{J}$. As a
consequence $\Delta _{\uparrow \downarrow }=2\omega _{\uparrow \downarrow }$
and the eigenvalues (\ref{eq25}) simplify as:
\begin{equation}
\begin{array}{cc}
\omega _{1}^{2}=\omega ^{2}-\omega _{\uparrow \downarrow }, & \omega
_{2}^{2}=\omega ^{2}+\omega _{\uparrow \downarrow }
\end{array}
.  \label{eq28}
\end{equation}
The operators $Q_{1},P_{1}$ and $Q_{2},P_{2}$ can be viewed as position and
momentum operators of two distinct fictitious particles, associated with the
modes $1$ and $2$, so that the Hamiltonian (\ref{eq24}) is a sum of two
harmonic oscillators. The corresponding Hilbert space is simply given by the
tensor product $\mathcal{E}_{\uparrow }\otimes \mathcal{E}_{\downarrow
}\equiv \mathcal{E}_{1}\otimes \mathcal{E}_{2}$ and two pairs of creation
and annihilation operators, one for each mode, can be introduced:
\begin{equation}
\begin{array}{cc}
a_{i}^{\dagger }=\frac{1}{\sqrt{2}}\left[ \sqrt{\frac{\omega _{i}}{\hbar }}%
Q_{i}-i\frac{P_{i}}{\sqrt{\omega _{i}\hbar }}\right] , & a_{i}=\frac{1}{%
\sqrt{2}}\left[ \sqrt{\frac{\omega _{i}}{\hbar }}Q_{i}+i\frac{P_{i}}{\sqrt{%
\omega _{i}\hbar }}\right] ,
\end{array}
\label{eq29}
\end{equation}
being $i=1,2$. Thus the stationary states of the full Hamiltonian (\ref{eq24}%
) are easily obtained:
\begin{equation}
\left| \varphi _{n,p}\right\rangle =\left| \varphi _{n}^{1}\right\rangle
\left| \varphi _{p}^{2}\right\rangle =\frac{1}{\sqrt{n!p!}}\left(
a_{1}^{\dagger }\right) ^{n}\left( a_{2}^{\dagger }\right) ^{p}\left|
\varphi _{0,0}\right\rangle ,  \label{eq30}
\end{equation}
together with the corresponding energies:
\begin{equation}
E_{n,p}=E_{n}^{1}+E_{p}^{2}=\left( n+\frac{1}{2}\right) \hbar \omega
_{1}+\left( p+\frac{1}{2}\right) \hbar \omega _{2}.  \label{eq31}
\end{equation}
Let us notice that this spectrum is not unbounded because an infinite number
of unphysical high energy states have been added. Thus a constraint has to
be included in order to satisfy the conditions $\langle a_{\uparrow
}^{\dagger }a_{\uparrow }\rangle \ll 2J^{\uparrow }$, $\langle a_{\downarrow
}^{\dagger }a_{\downarrow }\rangle \ll 2J^{\downarrow }$, required for the
validity of the linearized Holstein-Primakoff approximation. Solving these
constraints will give limits to the value of $n$ and $p$ and a finite
dimensional Hilbert space will be recovered.

We are interested in the time evolution of the mean values of the
observables $J_{x}^{\uparrow }$, $J_{x}^{\downarrow }$, that is the
population imbalance between the left and right well of the potential of
each pseudospin species. In order to carry out such a study and to impose
the correct initial conditions it is much more convenient to start from the
Heisenberg equations of motion for the observables $Q_{1}$, $Q_{2}$, $P_{1}$%
, $P_{2}$:
\begin{eqnarray}
\frac{d}{dt}\left\langle Q_{i}\right\rangle &=&\frac{1}{i\hbar }\left\langle %
\left[ Q_{i},H_{2BJJ}\right] \right\rangle =\left\langle P_{i}\right\rangle ,
\label{eq32} \\
\frac{d}{dt}\left\langle P_{i}\right\rangle &=&\frac{1}{i\hbar }\left\langle %
\left[ P_{i},H_{2BJJ}\right] \right\rangle =-\omega _{i}^{2}\left\langle
Q_{i}\right\rangle ,  \label{eq33}
\end{eqnarray}
which give rise to the following time evolution:
\begin{eqnarray}
\left\langle Q_{i}\right\rangle \left( t\right) &=&\left\langle
Q_{i}\right\rangle \left( 0\right) \cos \omega _{i}t+\frac{\left\langle
P_{i}\right\rangle \left( 0\right) }{\omega _{i}}\sin \omega _{i}t,
\label{eq34} \\
\left\langle P_{i}\right\rangle \left( t\right) &=&\left\langle
P_{i}\right\rangle \left( 0\right) \cos \omega _{i}t-\omega _{i}\left\langle
Q_{i}\right\rangle \left( 0\right) \sin \omega _{i}t.  \label{eq35}
\end{eqnarray}
We are ready to express $J_{x}^{\uparrow }$, $J_{x}^{\downarrow }$ in terms
of $Q_{1}$, $Q_{2}$, $P_{1}$, $P_{2}$ by means of Eqs. (\ref{eq20}), (\ref
{eq21}), (\ref{eq26}), (\ref{eq27}) and take their averages; in this way the
initial conditions $\left\langle J_{y}^{\uparrow }\right\rangle \left(
0\right) $, $\left\langle J_{y}^{\downarrow }\right\rangle \left( 0\right) $%
, $\left\langle J_{x}^{\uparrow }\right\rangle \left( 0\right) $, $%
\left\langle J_{x}^{\downarrow }\right\rangle \left( 0\right) $ are well
known. The final result is:
\begin{eqnarray}
\left\langle J_{x}^{\uparrow }\right\rangle \left( t\right) &=&\frac{%
a^{^{\prime }}\left\langle Q_{2}\right\rangle \left( t\right) -a^{^{\prime
\prime }}\left\langle Q_{1}\right\rangle \left( t\right) }{\left[ \frac{%
a^{^{\prime \prime }}b^{^{\prime }}}{\sqrt{-J^{\uparrow }J_{\uparrow
\uparrow }}}-\frac{a^{^{\prime }}b^{^{\prime \prime }}}{\sqrt{-J^{\downarrow
}J_{\downarrow \downarrow }}}\right] },  \label{eq36} \\
\left\langle J_{x}^{\downarrow }\right\rangle \left( t\right) &=&\frac{%
b^{^{\prime }}\frac{\sqrt{-J^{\downarrow }J_{\downarrow \downarrow }}}{\sqrt{%
-J^{\uparrow }J_{\uparrow \uparrow }}}\left\langle Q_{2}\right\rangle \left(
t\right) -b^{^{\prime \prime }}\left\langle Q_{1}\right\rangle \left(
t\right) }{\left[ \frac{a^{^{\prime \prime }}b^{^{\prime }}}{\sqrt{%
-J^{\uparrow }J_{\uparrow \uparrow }}}-\frac{a^{^{\prime }}b^{^{\prime
\prime }}}{\sqrt{-J^{\downarrow }J_{\downarrow \downarrow }}}\right] },
\label{eq37}
\end{eqnarray}
where the coefficients $a^{^{\prime }},b^{^{\prime }},a^{^{\prime \prime
}},b^{^{\prime \prime }}$ are defined as follows:
\begin{eqnarray}
a^{^{\prime }} &=&\frac{2\omega _{\uparrow \downarrow }}{\sqrt{4\omega
_{\uparrow \downarrow }^{2}+\left[ \left( \omega _{\downarrow }^{2}-\omega
_{\uparrow }^{2}\right) +\sqrt{\left( \omega _{\uparrow }^{2}-\omega
_{\downarrow }^{2}\right) ^{2}+4\omega _{\uparrow \downarrow }^{2}}\right]
^{2}}},  \label{eq38} \\
b^{^{\prime }} &=&\frac{\left( \omega _{\downarrow }^{2}-\omega _{\uparrow
}^{2}\right) +\sqrt{\left( \omega _{\uparrow }^{2}-\omega _{\downarrow
}^{2}\right) ^{2}+4\omega _{\uparrow \downarrow }^{2}}}{\sqrt{4\omega
_{\uparrow \downarrow }^{2}+\left[ \left( \omega _{\downarrow }^{2}-\omega
_{\uparrow }^{2}\right) +\sqrt{\left( \omega _{\uparrow }^{2}-\omega
_{\downarrow }^{2}\right) ^{2}+4\omega _{\uparrow \downarrow }^{2}}\right]
^{2}}},  \label{eq39} \\
a^{^{\prime \prime }} &=&\frac{2\omega _{\uparrow \downarrow }}{\sqrt{%
4\omega _{\uparrow \downarrow }^{2}+\left[ \left( \omega _{\downarrow
}^{2}-\omega _{\uparrow }^{2}\right) -\sqrt{\left( \omega _{\uparrow
}^{2}-\omega _{\downarrow }^{2}\right) ^{2}+4\omega _{\uparrow \downarrow
}^{2}}\right] ^{2}}},  \label{eq40} \\
b^{^{\prime \prime }} &=&\frac{\left( \omega _{\downarrow }^{2}-\omega
_{\uparrow }^{2}\right) -\sqrt{\left( \omega _{\uparrow }^{2}-\omega
_{\downarrow }^{2}\right) ^{2}+4\omega _{\uparrow \downarrow }^{2}}}{\sqrt{%
4\omega _{\uparrow \downarrow }^{2}+\left[ \left( \omega _{\downarrow
}^{2}-\omega _{\uparrow }^{2}\right) -\sqrt{\left( \omega _{\uparrow
}^{2}-\omega _{\downarrow }^{2}\right) ^{2}+4\omega _{\uparrow \downarrow
}^{2}}\right] ^{2}}}.  \label{eq41}
\end{eqnarray}
In this limiting case we could impose initial conditions featuring a small
imbalance between the two wells such as, for instance, $\left\langle
J_{x}^{\uparrow }\right\rangle \left( 0\right) =\pm 1$, $\left\langle
J_{x}^{\downarrow }\right\rangle \left( 0\right) =\pm 1$, $\left\langle
J_{y}^{\uparrow }\right\rangle \left( 0\right) =0$, $\left\langle
J_{y}^{\downarrow }\right\rangle \left( 0\right) =0$, and vary the physical
parameters $J_{\uparrow \uparrow }$, $J_{\downarrow \downarrow }$, $%
\overline{g}_{\uparrow \uparrow }$, $\overline{g}_{\downarrow \downarrow }$,
$\overline{g}_{\uparrow \downarrow }$ in a wide range. The results show up a
coherent tunneling of each pseudospin species between the two wells for
negative $\overline{g}_{\uparrow \downarrow }$ and a phase separation
instability upon increasing $\overline{g}_{\uparrow \downarrow }$ above a
critical positive value. Furthermore within the above\ analysis only
short-time scale effects are reliable. This phenomenology coincides with our
previous results for a binary mixture of BECs, so we refer to Ref. \cite
{noi2} for further details.

\subsection{Strong Raman coupling regime}

This limiting case corresponds to the physical situation $\overline{\Omega }%
>>\left| J_{\uparrow \uparrow }\right| ,\left| J_{\downarrow \downarrow
}\right| $ \ while $\overline{\Omega }\lesssim E_{L}$, so that atomic
interwell tunneling can be neglected and the relevant phenomenology is due
to the internal dynamics in each single well. As a consequence the total
number of particles is conserved for the left as well as the right well, $%
N_{L}=N_{L\uparrow }+N_{L\downarrow }$ and $N_{R}=N_{R\uparrow
}+N_{R\downarrow }$, respectively.

In this limit Hamiltonian (\ref{eq12}) reduces to:
\begin{equation}
\mathcal{H}=\mathcal{H}_{L}+\mathcal{H}_{R},  \label{eq42}
\end{equation}
where
\begin{equation}
\mathcal{H}_{i}=\frac{\overline{\Omega }}{2}\left( a_{i\uparrow }^{\dagger
}a_{i\downarrow }+a_{i\downarrow }^{\dagger }a_{i\uparrow }\right) +\frac{%
\delta }{2}\left( a_{i\uparrow }^{\dagger }a_{i\uparrow }-a_{i\downarrow
}^{\dagger }a_{i\downarrow }\right) +\frac{1}{2}\overline{g}_{\uparrow
\uparrow }a_{i\uparrow }^{\dagger }a_{i\uparrow }^{\dagger }a_{i\uparrow
}a_{i\uparrow }+\frac{1}{2}\overline{g}_{\downarrow \downarrow
}a_{i\downarrow }^{\dagger }a_{i\downarrow }^{\dagger }a_{i\downarrow
}a_{i\downarrow }+\overline{g}_{\uparrow \downarrow }a_{i\uparrow }^{\dagger
}a_{i\downarrow }^{\dagger }a_{i\uparrow }a_{i\downarrow },  \label{eq43}
\end{equation}
being $i=L,R$. \ Clearly we deal with two independent internal Josephson
effects, one for each well, and the solution proceeds exactly as in the
previous regime, through a mapping to a $SU(2)$ problem followed by a
linearized Holstein-Primakoff transformation \cite{hp1,hp2}. Thus, by
performing the same steps as in the previous subsection, Hamiltonian (\ref
{eq42}) can be expressed as a sum of two independent harmonic oscillators,
in the left and in the right well of the potential, respectively:
\begin{equation}
\mathcal{H}\simeq \frac{1}{2}\left[ \omega _{L}^{2}\overline{Q}%
_{L}^{2}+P_{L}^{2}+\omega _{R}^{2}\overline{Q}_{R}^{2}+P_{R}^{2}\right] ,
\label{eq44}
\end{equation}
where $\omega _{i}^{2}=1-J^{i}\overline{\Omega }\left( \overline{g}%
_{\uparrow \uparrow }+\overline{g}_{\downarrow \downarrow }-2\overline{g}%
_{\uparrow \downarrow }\right) $, $\overline{Q}_{i}=Q_{i}+\frac{\left[
\left( N_{i}-1\right) \left( \overline{g}_{\downarrow \downarrow }-\overline{%
g}_{\uparrow \uparrow }\right) +2\delta \right] \sqrt{J^{i}}\sqrt{-\overline{%
\Omega }}}{2\omega _{i}^{2}}$, $i=L,R$ and we have dropped constant terms $%
C.st=\overline{\Omega }\left( J^{L}-\frac{1}{2}\right) +\overline{\Omega }%
\left( J^{R}-\frac{1}{2}\right) +\frac{1}{8}\frac{J^{L}\overline{\Omega }%
\left[ \left( N_{L}-1\right) \left( \overline{g}_{\downarrow \downarrow }-%
\overline{g}_{\uparrow \uparrow }\right) +2\delta \right] ^{2}}{\omega
_{L}^{2}}+\frac{1}{8}\frac{J^{R}\overline{\Omega }\left[ \left(
N_{R}-1\right) \left( \overline{g}_{\downarrow \downarrow }-\overline{g}%
_{\uparrow \uparrow }\right) +2\delta \right] ^{2}}{\omega _{R}^{2}}$.

Finally, the time evolution of the population imbalance between the two
pseudospin states within each well is:
\begin{equation}
\left\langle J_{x}^{L}\right\rangle \left( t\right) =\sqrt{J^{L}}\sqrt{-%
\overline{\Omega }}\left\langle \overline{Q}_{L}\right\rangle \left(
t\right) -\frac{\overline{\Omega }J^{L}\left[ \left( N_{L}-1\right) \left(
\overline{g}_{\downarrow \downarrow }-\overline{g}_{\uparrow \uparrow
}\right) +2\delta \right] }{2\omega _{L}^{2}},  \label{eq45}
\end{equation}
\begin{equation}
\left\langle J_{x}^{R}\right\rangle \left( t\right) =\sqrt{J^{L}}\sqrt{-%
\overline{\Omega }}\left\langle \overline{Q}_{R}\right\rangle \left(
t\right) -\frac{\overline{\Omega }J^{R}\left[ \left( N_{R}-1\right) \left(
\overline{g}_{\downarrow \downarrow }-\overline{g}_{\uparrow \uparrow
}\right) +2\delta \right] }{2\omega _{R}^{2}},  \label{eq46}
\end{equation}
where
\begin{equation}
\begin{array}{cc}
\left\langle \overline{Q}_{i}\right\rangle \left( t\right) =\left\langle
\overline{Q}_{i}\right\rangle \left( 0\right) \cos \omega _{i}t+\frac{%
\left\langle P_{i}\right\rangle \left( 0\right) }{\omega _{i}}\sin \omega
_{i}t, & i=L,R,
\end{array}
\label{eq47}
\end{equation}
and suitable initial conditions $\left\langle J_{x}^{L}\right\rangle \left(
0\right) $, $\left\langle J_{x}^{R}\right\rangle \left( 0\right) $ have to
be imposed. The resulting Josephson oscillations are entirely induced by the
Raman coupling and take place within the spin space.

\section{Quantum dynamics within the intermediate regime}

In this Section we deal with the intermediate regime, where $\overline{%
\Omega }$ and $J_{\uparrow \uparrow }$, $J_{\downarrow \downarrow }$ equally
contribute to the dynamics and the problem is much more involved. In
general, one cannot obtain a closed analytical solution and has to resort to
numerical calculations. Here we concentrate on some particular cases
amenable to an analytical solution while the full numerical calculation will
be the subject of a future publication \cite{noiF}.

\subsection{The noninteracting case}

In order to gain some insight into the phenomenology we start with the
strong tunneling regime and neglect collisional interactions, i. e. we put $%
\overline{g}_{\uparrow \uparrow }=\overline{g}_{\downarrow \downarrow }=%
\overline{g}_{\uparrow \downarrow }$. We also make a further simplifying
assumption, i. e. $J_{\uparrow \uparrow }=J_{\downarrow \downarrow }=%
\overline{J}$, so that Hamiltonian (\ref{eq12}) reduces to
\begin{eqnarray}
\mathcal{H} &=&\overline{J}\left( a_{L\uparrow }^{\dagger }a_{R\uparrow
}+a_{R\uparrow }^{\dagger }a_{L\uparrow }+a_{L\downarrow }^{\dagger
}a_{R\downarrow }+a_{R\downarrow }^{\dagger }a_{L\downarrow }\right) +\frac{%
\overline{\Omega }}{2}\left( a_{L\uparrow }^{\dagger }a_{L\downarrow
}+a_{L\downarrow }^{\dagger }a_{L\uparrow }+a_{R\uparrow }^{\dagger
}a_{R\downarrow }+a_{R\downarrow }^{\dagger }a_{R\uparrow }\right)  \nonumber
\\
&&+\frac{\delta }{2}\left( a_{L\uparrow }^{\dagger }a_{L\uparrow
}+a_{R\uparrow }^{\dagger }a_{R\uparrow }-a_{L\downarrow }^{\dagger
}a_{L\downarrow }-a_{R\downarrow }^{\dagger }a_{R\downarrow }\right) .
\label{eq48}
\end{eqnarray}
By taking a closer look to the above expression we recognize a quadratic
Hamiltonian which can be promptly diagonalized by introducing the following
transformation:
\begin{eqnarray}
c_{1} &=&\frac{\overline{\Omega }}{\sqrt{2\overline{\Omega }^{2}+2\left(
\delta +\overline{\Omega }_{\delta }\right) ^{2}}}\left[ \left( \frac{\delta
+\overline{\Omega }_{\delta }}{\overline{\Omega }}\right) a_{L\uparrow
}+a_{L\downarrow }+\left( \frac{\delta +\overline{\Omega }_{\delta }}{%
\overline{\Omega }}\right) a_{R\uparrow }+a_{R\downarrow }\right] ,
\label{eq49} \\
c_{2} &=&\frac{\overline{\Omega }}{\sqrt{2\overline{\Omega }^{2}+2\left(
\delta -\overline{\Omega }_{\delta }\right) ^{2}}}\left[ \left( \frac{%
-\delta +\overline{\Omega }_{\delta }}{\overline{\Omega }}\right)
a_{L\uparrow }-a_{L\downarrow }+\left( \frac{\delta -\overline{\Omega }%
_{\delta }}{\overline{\Omega }}\right) a_{R\uparrow }+a_{R\downarrow }\right]
,  \label{eq50} \\
d_{1} &=&\frac{\overline{\Omega }}{\sqrt{2\overline{\Omega }^{2}+2\left(
\delta +\overline{\Omega }_{\delta }\right) ^{2}}}\left[ -\left( \frac{%
\delta +\overline{\Omega }_{\delta }}{\overline{\Omega }}\right)
a_{L\uparrow }-a_{L\downarrow }+\left( \frac{\delta +\overline{\Omega }%
_{\delta }}{\overline{\Omega }}\right) a_{R\uparrow }+a_{R\downarrow }\right]
,  \label{eq51} \\
d_{2} &=&\frac{\overline{\Omega }}{\sqrt{2\overline{\Omega }^{2}+2\left(
\delta -\overline{\Omega }_{\delta }\right) ^{2}}}\left[ \left( \frac{\delta
-\overline{\Omega }_{\delta }}{\overline{\Omega }}\right) a_{L\uparrow
}+a_{L\downarrow }+\left( \frac{\delta -\overline{\Omega }_{\delta }}{%
\overline{\Omega }}\right) a_{R\uparrow }+a_{R\downarrow }\right] ,
\label{eq52}
\end{eqnarray}
where $\overline{\Omega }_{\delta }=\sqrt{\delta ^{2}+\overline{\Omega }^{2}}
$ and the usual commutation relations hold: $\left[ c_{i},c_{j}^{\dagger }%
\right] =\delta _{ij}$, $\left[ d_{i},d_{j}^{\dagger }\right] =\delta _{ij}$%
. As a consequence, Hamiltonian (\ref{eq48}) can be cast in the simple form
\begin{equation}
\mathcal{H}=\left( \frac{\overline{\Omega }_{\delta }}{2}+\overline{J}%
\right) \left( c_{1}^{\dagger }c_{1}-c_{2}^{\dagger }c_{2}\right) +\left(
\frac{\overline{\Omega }_{\delta }}{2}-\overline{J}\right) \left(
d_{1}^{\dagger }d_{1}-d_{2}^{\dagger }d_{2}\right)  \label{eq53}
\end{equation}
and the conservation relation $N=c_{1}^{\dagger }c_{1}+c_{2}^{\dagger
}c_{2}+d_{1}^{\dagger }d_{1}+d_{2}^{\dagger }d_{2}$ is satisfied. Let us
notice that, in the new basis, the quantum dynamics is characterized by two
frequencies: $\omega _{c}=\frac{\overline{\Omega }_{\delta }}{2}+\overline{J}
$ and $\omega _{d}=\frac{\overline{\Omega }_{\delta }}{2}-\overline{J}$.
This will appear more clearly in the time evolution of particle imbalances
between the two wells as well between the two spin states, which we now
study. Let us start by the assumption that at time $t=0$ all $N$ atoms have
pseudospin down and lie in the left well. The corresponding initial
condition reads as:
\begin{equation}
\left| \psi \left( t=0\right) \right\rangle =\frac{1}{\sqrt{N!}}\left(
a_{L\downarrow }^{\dagger }\right) ^{N}\left| 0\right\rangle =\frac{1}{2^{N}%
\sqrt{N!}}\left( \sqrt{\frac{\overline{\Omega }_{\delta }-\delta }{\overline{%
\Omega }_{\delta }}}c_{1}^{\dagger }-\sqrt{\frac{\overline{\Omega }_{\delta
}+\delta }{\overline{\Omega }_{\delta }}}c_{2}^{\dagger }-\sqrt{\frac{%
\overline{\Omega }_{\delta }-\delta }{\overline{\Omega }_{\delta }}}%
d_{1}^{\dagger }+\sqrt{\frac{\overline{\Omega }_{\delta }+\delta }{\overline{%
\Omega }_{\delta }}}d_{2}^{\dagger }\right) ^{N}\left| 0\right\rangle .
\label{eq54}
\end{equation}
Then let us switch on tunneling as well as Raman coupling terms and
determine the dynamics at any time $t$. By performing the unitary
transformation
\begin{eqnarray}
&&e^{-i\mathcal{H}t}\left( \sqrt{\frac{\overline{\Omega }_{\delta }-\delta }{%
\overline{\Omega }_{\delta }}}c_{1}^{\dagger }-\sqrt{\frac{\overline{\Omega }%
_{\delta }+\delta }{\overline{\Omega }_{\delta }}}c_{2}^{\dagger }-\sqrt{%
\frac{\overline{\Omega }_{\delta }-\delta }{\overline{\Omega }_{\delta }}}%
d_{1}^{\dagger }+\sqrt{\frac{\overline{\Omega }_{\delta }+\delta }{\overline{%
\Omega }_{\delta }}}d_{2}^{\dagger }\right) e^{i\mathcal{H}t}  \nonumber \\
&\equiv &e^{-i\omega _{c}t}\sqrt{\frac{\overline{\Omega }_{\delta }-\delta }{%
\overline{\Omega }_{\delta }}}c_{1}^{\dagger }-e^{i\omega _{c}t}\sqrt{\frac{%
\overline{\Omega }_{\delta }+\delta }{\overline{\Omega }_{\delta }}}%
c_{2}^{\dagger }-e^{-i\omega _{d}t}\sqrt{\frac{\overline{\Omega }_{\delta
}-\delta }{\overline{\Omega }_{\delta }}}d_{1}^{\dagger }+e^{i\omega _{d}t}%
\sqrt{\frac{\overline{\Omega }_{\delta }+\delta }{\overline{\Omega }_{\delta
}}}d_{2}^{\dagger },  \label{eq55}
\end{eqnarray}
we get the wave function at time $t$:
\begin{equation}
\left| \psi \left( t\right) \right\rangle =\frac{1}{2^{N}\sqrt{N!}}\left(
e^{-i\omega _{c}t}\sqrt{\frac{\overline{\Omega }_{\delta }-\delta }{%
\overline{\Omega }_{\delta }}}c_{1}^{\dagger }-e^{i\omega _{c}t}\sqrt{\frac{%
\overline{\Omega }_{\delta }+\delta }{\overline{\Omega }_{\delta }}}%
c_{2}^{\dagger }-e^{-i\omega _{d}t}\sqrt{\frac{\overline{\Omega }_{\delta
}-\delta }{\overline{\Omega }_{\delta }}}d_{1}^{\dagger }+e^{i\omega _{d}t}%
\sqrt{\frac{\overline{\Omega }_{\delta }+\delta }{\overline{\Omega }_{\delta
}}}d_{2}^{\dagger }\right) ^{N}\left| 0\right\rangle ,  \label{eq56}
\end{equation}
which, by substituting Eqs. (\ref{eq49})-(\ref{eq52}), takes the final form
\begin{equation}
\left| \psi \left( t\right) \right\rangle =\frac{1}{\sqrt{N!}}\left(
G_{L\uparrow }\left( t\right) a_{L\uparrow }^{\dagger }+G_{L\downarrow
}\left( t\right) a_{L\downarrow }^{\dagger }+G_{R\uparrow }\left( t\right)
a_{R\uparrow }^{\dagger }+G_{R\downarrow }\left( t\right) a_{R\downarrow
}^{\dagger }\right) ^{N}\left| 0\right\rangle .  \label{eq57}
\end{equation}
Here the functions $G_{j\sigma }\left( t\right) $, $j=L,R$, $\sigma
=\uparrow ,\downarrow $ are defined as:
\begin{eqnarray}
G_{L\uparrow }\left( t\right) &=&-i\frac{\overline{\Omega }}{\overline{%
\Omega }_{\delta }}\frac{\left[ \sin \left( \omega _{c}t\right) +\sin \left(
\omega _{d}t\right) \right] }{2},  \label{eq58} \\
G_{L\downarrow }\left( t\right) &=&\frac{\left[ \cos \left( \omega
_{c}t\right) +\cos \left( \omega _{d}t\right) \right] }{2}+i\frac{\delta }{%
\overline{\Omega }_{\delta }}\frac{\left[ \sin \left( \omega _{c}t\right)
+\sin \left( \omega _{d}t\right) \right] }{2},  \label{eq59} \\
G_{R\uparrow }\left( t\right) &=&\frac{\overline{\Omega }}{\overline{\Omega }%
_{\delta }}\frac{\left[ \cos \left( \omega _{c}t\right) -\cos \left( \omega
_{d}t\right) \right] }{2},  \label{eq60} \\
G_{R\downarrow }\left( t\right) &=&-i\frac{\left[ \sin \left( \omega
_{c}t\right) -\sin \left( \omega _{d}t\right) \right] }{2}+\frac{\delta }{%
\overline{\Omega }_{\delta }}\frac{\left[ \cos \left( \omega _{c}t\right)
-\cos \left( \omega _{d}t\right) \right] }{2},  \label{eq61}
\end{eqnarray}
and are characterized by the two frequencies $\omega _{c}$ and $\omega _{d}$.

In order to calculate the particle imbalance between the two wells and the
two pseudospin states let us introduce the following quantity:
\begin{equation}
g_{j,\sigma ;k,\sigma ^{\prime }}\left( t\right) =\frac{1}{N}\left\langle
\psi \left( t\right) \right| a_{j\sigma }^{\dagger }a_{k\sigma ^{\prime
}}\left| \psi \left( t\right) \right\rangle =G_{j\sigma }^{\ast }\left(
t\right) G_{k\sigma ^{\prime }}\left( t\right) ,  \label{eq62}
\end{equation}
from which the fraction of pseudospin $\sigma $ in the $j$ well is easily
obtained:
\begin{equation}
n_{j,\sigma }\left( t\right) =g_{j,\sigma ;j,\sigma }\left( t\right) =\left|
G_{j\sigma }\left( t\right) \right| ^{2}.  \label{eq63}
\end{equation}
The required particle imbalance between left and right well, which gives
rise to external Josephson oscillations, reads:
\begin{equation}
\rho _{\sigma }\left( t\right) =n_{L,\sigma }\left( t\right) -n_{R,\sigma
}\left( t\right) =\left| G_{L\sigma }\left( t\right) \right| ^{2}-\left|
G_{R\sigma }\left( t\right) \right| ^{2},  \label{eq64}
\end{equation}
while internal Josephson oscillations are governed by the pseudospin
imbalance:
\begin{equation}
\rho _{j}\left( t\right) =n_{j,\downarrow }\left( t\right) -n_{j,\uparrow
}\left( t\right) =\left| G_{j\downarrow }\left( t\right) \right| ^{2}-\left|
G_{j\uparrow }\left( t\right) \right| ^{2}.  \label{eq65}
\end{equation}
The net result is an interesting interplay between external and internal
Josephson effects.

Indeed we can calculate the population imbalance between the two wells
\begin{equation}
D_{LR}\left( t\right) =\rho _{\uparrow }\left( t\right) +\rho _{\downarrow
}\left( t\right) =\left| G_{L\uparrow }\left( t\right) \right| ^{2}-\left|
G_{R\uparrow }\left( t\right) \right| ^{2}+\left| G_{L\downarrow }\left(
t\right) \right| ^{2}-\left| G_{R\downarrow }\left( t\right) \right| ^{2},
\label{eq65a}
\end{equation}
the magnetization
\begin{equation}
M_{LR}\left( t\right) =\rho _{\uparrow }\left( t\right) -\rho _{\downarrow
}\left( t\right) =\left| G_{L\uparrow }\left( t\right) \right| ^{2}-\left|
G_{R\uparrow }\left( t\right) \right| ^{2}-\left| G_{L\downarrow }\left(
t\right) \right| ^{2}+\left| G_{R\downarrow }\left( t\right) \right| ^{2},
\label{eq65b}
\end{equation}
and the pseudospin imbalance
\begin{equation}
D_{\uparrow \downarrow }\left( t\right) =\rho _{L}\left( t\right) +\rho
_{R}\left( t\right) =\left| G_{L\downarrow }\left( t\right) \right|
^{2}-\left| G_{L\uparrow }\left( t\right) \right| ^{2}+\left| G_{R\downarrow
}\left( t\right) \right| ^{2}-\left| G_{R\uparrow }\left( t\right) \right|
^{2}.  \label{eq65c}
\end{equation}

In order to show the general features of the above results we report in Figs. 1, 2 and 3 the behaviours of Eqs. (\ref{eq65a})-(\ref{eq65c}) as a function of time for the following choice of parameters: $\overline{\Omega }=0.1 E_{L}$, $\overline{J }=-0.1 E_{L}$ and $\delta=0.01 E_{L}$. Such a choice is in close agreement with the experimental parameters by the NIST group \cite{nist2}. Indeed the results show up coherent Rabi type oscillations both for the population imbalance between the two wells (external Josephson tunneling) and the pseudospin imbalance between up and down states (internal Josephson tunneling), even if with different frequencies. Conversely the magnetization $M_{LR}$ exhibits complicated quasiperiodic features. Finally, by varying parameters over a broad range between weak ($\overline{\Omega }<<\left| \overline{J }\right| $) and strong ($\overline{\Omega }>>\left| \overline{J }\right| $) Raman coupling similar behaviours have been found. We will show in the next Subsection how this picture modifies in the presence of nonlinear interactions.

\begin{figure}[tbph]
\centering
\includegraphics[scale=0.8]{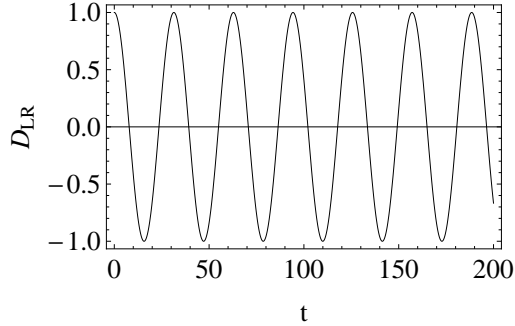}\\
\caption{
Behavior of the population imbalance $D_{LR}$ between the two wells for $\overline{\Omega }=0.1$, $\overline{J }=-0.1$ and $\delta=0.01$ (units of $E_{L}$). The time is expressed in units of $\frac{\hbar}{E_{L}}$.}\label{fig:fig1}
\end{figure}
\begin{figure}[tbph]
\centering
\includegraphics[scale=0.8]{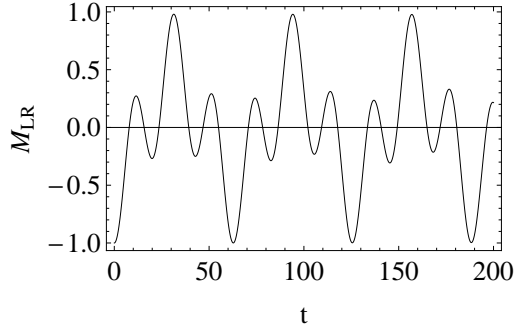}\\
\caption{
Behavior of the magnetization $M_{LR}$ between the two wells for $\overline{\Omega }=0.1$, $\overline{J }=-0.1$ and $\delta=0.01$ (units of $E_{L}$). The time is expressed in units of $\frac{\hbar}{E_{L}}$.}\label{fig:fig2}
\end{figure}
\begin{figure}[tbph]
\centering
\includegraphics[scale=0.8]{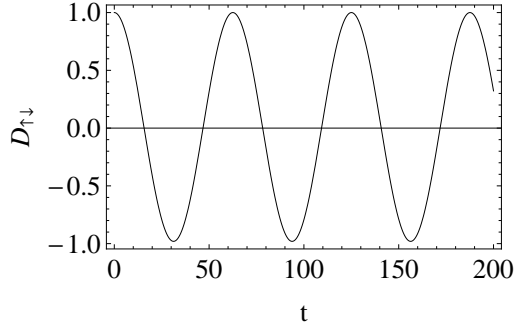}\\
\caption{
Behavior of the pseudospin imbalance $D_{\uparrow \downarrow }$ between the two wells for $\overline{\Omega }=0.1$, $\overline{J }=-0.1$ and $\delta=0.01$ (units of $E_{L}$). The time is expressed in units of $\frac{\hbar}{E_{L}}$.}\label{fig:fig3}
\end{figure}

\subsection{Weak collisional interaction limit}

In this Subsection we will consider a particular case, which admits a simple
analytical solution. Let us put $\overline{g}_{\uparrow \uparrow }=\overline{%
g}_{\downarrow \downarrow }=0$ and $\overline{g}_{\uparrow \downarrow }=%
\overline{g}\neq 0$ while retaining $J_{\uparrow \uparrow }=J_{\downarrow
\downarrow }=\overline{J}$, so that Hamiltonian (\ref{eq12}) reduces to
\begin{eqnarray}
\mathcal{H} &=&\overline{J}\left( a_{L\uparrow }^{\dagger }a_{R\uparrow
}+a_{R\uparrow }^{\dagger }a_{L\uparrow }+a_{L\downarrow }^{\dagger
}a_{R\downarrow }+a_{R\downarrow }^{\dagger }a_{L\downarrow }\right) +\frac{%
\overline{\Omega }}{2}\left( a_{L\uparrow }^{\dagger }a_{L\downarrow
}+a_{L\downarrow }^{\dagger }a_{L\uparrow }+a_{R\uparrow }^{\dagger
}a_{R\downarrow }+a_{R\downarrow }^{\dagger }a_{R\uparrow }\right)  \nonumber
\\
&&+\frac{\delta }{2}\left( a_{L\uparrow }^{\dagger }a_{L\uparrow
}+a_{R\uparrow }^{\dagger }a_{R\uparrow }-a_{L\downarrow }^{\dagger
}a_{L\downarrow }-a_{R\downarrow }^{\dagger }a_{R\downarrow }\right) +%
\overline{g}\left( a_{L\uparrow }^{\dagger }a_{L\downarrow }^{\dagger
}a_{L\uparrow }a_{L\downarrow }+a_{R\uparrow }^{\dagger }a_{R\downarrow
}^{\dagger }a_{R\uparrow }a_{R\downarrow }\right) .  \label{eq66}
\end{eqnarray}
The solution of previous noninteracting case leads us to perform the
following transformation:
\begin{eqnarray}
a_{L\uparrow } &=&\frac{\overline{\Omega }}{2\overline{\Omega }_{\delta }}%
\left[ \sqrt{\frac{\overline{\Omega }_{\delta }}{\overline{\Omega }_{\delta
}-\delta }}e^{i2\omega _{c}t}c_{1}+\sqrt{\frac{\overline{\Omega }_{\delta }}{%
\overline{\Omega }_{\delta }+\delta }}e^{-i2\omega _{c}t}c_{2}-\sqrt{\frac{%
\overline{\Omega }_{\delta }}{\overline{\Omega }_{\delta }-\delta }}%
e^{i2\omega _{d}t}d_{1}-\sqrt{\frac{\overline{\Omega }_{\delta }}{\overline{%
\Omega }_{\delta }+\delta }}e^{-i2\omega _{d}t}d_{2}\right] ,  \label{eq67}
\\
a_{L\downarrow } &=&\frac{1}{2}\left[ \sqrt{\frac{\overline{\Omega }_{\delta
}-\delta }{\overline{\Omega }_{\delta }}}e^{i2\omega _{c}t}c_{1}-\sqrt{\frac{%
\overline{\Omega }_{\delta }+\delta }{\overline{\Omega }_{\delta }}}%
e^{-i2\omega _{c}t}c_{2}-\sqrt{\frac{\overline{\Omega }_{\delta }-\delta }{%
\overline{\Omega }_{\delta }}}e^{i2\omega _{d}t}d_{1}+\sqrt{\frac{\overline{%
\Omega }_{\delta }+\delta }{\overline{\Omega }_{\delta }}}e^{-i2\omega
_{d}t}d_{2}\right] ,  \label{eq68} \\
a_{R\uparrow } &=&\frac{\overline{\Omega }}{2\overline{\Omega }_{\delta }}%
\left[ \sqrt{\frac{\overline{\Omega }_{\delta }}{\overline{\Omega }_{\delta
}-\delta }}e^{i2\omega _{c}t}c_{1}-\sqrt{\frac{\overline{\Omega }_{\delta }}{%
\overline{\Omega }_{\delta }+\delta }}e^{-i2\omega _{c}t}c_{2}+\sqrt{\frac{%
\overline{\Omega }_{\delta }}{\overline{\Omega }_{\delta }-\delta }}%
e^{i2\omega _{d}t}d_{1}-\sqrt{\frac{\overline{\Omega }_{\delta }}{\overline{%
\Omega }_{\delta }+\delta }}e^{-i2\omega _{d}t}d_{2}\right] ,  \label{eq69}
\\
a_{R\downarrow } &=&\frac{1}{2}\left[ \sqrt{\frac{\overline{\Omega }_{\delta
}-\delta }{\overline{\Omega }_{\delta }}}e^{i2\omega _{c}t}c_{1}+\sqrt{\frac{%
\overline{\Omega }_{\delta }+\delta }{\overline{\Omega }_{\delta }}}%
e^{-i2\omega _{c}t}c_{2}+\sqrt{\frac{\overline{\Omega }_{\delta }-\delta }{%
\overline{\Omega }_{\delta }}}e^{i2\omega _{d}t}d_{1}+\sqrt{\frac{\overline{%
\Omega }_{\delta }+\delta }{\overline{\Omega }_{\delta }}}e^{-i2\omega
_{d}t}d_{2}\right] ,  \label{eq70}
\end{eqnarray}
where the usual commutation relations hold: $\left[ c_{i},c_{j}^{\dagger }%
\right] =\delta _{ij}$, $\left[ d_{i},d_{j}^{\dagger }\right] =\delta _{ij}$%
. Rotating wave approximation \cite{rw1}\cite{rw2} allows us to drop fast
oscillating terms while retaining resonant terms. Proceeding along this line
and considering a parameters regime such that $\delta <<\overline{\Omega }$
Hamiltonian (\ref{eq66}) can be cast in the simple form:
\begin{eqnarray}
\mathcal{H} &=&\left( \frac{\overline{\Omega }_{\delta }}{2}+\overline{J}%
\right) \left( c_{1}^{\dagger }c_{1}-c_{2}^{\dagger }c_{2}\right) +\left(
\frac{\overline{\Omega }_{\delta }}{2}-\overline{J}\right) \left(
d_{1}^{\dagger }d_{1}-d_{2}^{\dagger }d_{2}\right) +\frac{\overline{g}}{8%
\overline{\Omega }_{\delta }^{2}}\left[ \overline{\Omega }^{2}c_{1}^{\dagger
}c_{1}^{\dagger }c_{1}c_{1}\right.  \nonumber \\
&&\left. +\overline{\Omega }^{2}c_{2}^{\dagger }c_{2}^{\dagger }c_{2}c_{2}+%
\overline{\Omega }^{2}d_{1}^{\dagger }d_{1}^{\dagger }d_{1}d_{1}+\overline{%
\Omega }^{2}d_{2}^{\dagger }d_{2}^{\dagger }d_{2}d_{2}+4\overline{\Omega }%
^{2}c_{1}^{\dagger }d_{1}^{\dagger }c_{1}d_{1}+4\overline{\Omega }%
^{2}c_{2}^{\dagger }d_{2}^{\dagger }c_{2}d_{2}\right.  \label{eq71} \\
&&\left. +4\delta ^{2}c_{1}^{\dagger }c_{2}^{\dagger }c_{1}c_{2}+4\delta
^{2}c_{1}^{\dagger }d_{2}^{\dagger }c_{1}d_{2}+4\delta ^{2}c_{2}^{\dagger
}d_{1}^{\dagger }c_{2}d_{1}+4\delta ^{2}d_{1}^{\dagger }d_{2}^{\dagger
}d_{1}c_{2}\right] .  \nonumber
\end{eqnarray}
This Hamiltonian is already diagonal in the Fock basis $\left|
mnpq\right\rangle $ and the corresponding energy eigenvalues are:
\begin{eqnarray}
E_{m,n,p,q} &=&\left( \frac{\overline{\Omega }_{\delta }}{2}+\overline{J}%
\right) \left( m-n\right) +\left( \frac{\overline{\Omega }_{\delta }}{2}-%
\overline{J}\right) \left( p-q\right)  \nonumber \\
&&+\frac{\overline{g}\overline{\Omega }^{2}}{8\overline{\Omega }_{\delta
}^{2}}\left( m^{2}-m+n^{2}-n+p^{2}-p+q^{2}-q+4mp+4nq\right)  \label{eq72} \\
&&+\frac{\overline{g}\delta ^{2}}{8\overline{\Omega }_{\delta }^{2}}\left(
4mn+4mq+4np+4pq\right) .  \nonumber
\end{eqnarray}
By choosing the same initial condition as in the previous Subsection (i. e.
at time $t=0$ all $N$ atoms have pseudospin down and lie in the left well)
the wavefunction at $t=0$ expressed in the Fock basis is:
\begin{equation}
\left| \psi \left( t=0\right) \right\rangle =\frac{1}{\sqrt{N!}}\left(
a_{L\downarrow }^{\dagger }\right) ^{N}\left| 0\right\rangle =\frac{1}{2^{N}%
\overline{\Omega }_{\delta }^{N/2}}\sum_{mnpq}\delta _{N,m+n+p+q}\frac{\sqrt{%
N!}A^{m}B^{n}C^{p}D^{q}}{\sqrt{m!n!p!q!}}\left| 0\right\rangle ,
\label{eq73}
\end{equation}
where $A=\sqrt{\overline{\Omega }_{\delta }-\delta }$, $B=-\sqrt{\overline{%
\Omega }_{\delta }+\delta }$, $C=-\sqrt{\overline{\Omega }_{\delta }-\delta }
$, $D=\sqrt{\overline{\Omega }_{\delta }+\delta }$. Thus, at time $t$ the
corresponding wavefunction reads:
\begin{eqnarray}
\left| \psi \left( t\right) \right\rangle &=&\frac{1}{2^{N}\overline{\Omega }%
_{\delta }^{N/2}}e^{-i\mathcal{H}t}\sum_{mnpq}\delta _{N,m+n+p+q}\frac{\sqrt{%
N!}A^{m}B^{n}C^{p}D^{q}}{\sqrt{m!n!p!q!}}\left| 0\right\rangle  \nonumber \\
&=&\frac{1}{2^{N}\overline{\Omega }_{\delta }^{N/2}}\sum_{mnpq}\delta
_{N,m+n+p+q}\frac{\sqrt{N!}A^{m}B^{n}C^{p}D^{q}}{\sqrt{m!n!p!q!}}%
e^{-iE_{m,n,p,q}t}\left| 0\right\rangle .  \label{eq74}
\end{eqnarray}
In order to calculate the fraction of pseudospin $\sigma $ in the $j$ well, $%
n_{j,\sigma }\left( t\right) $, as the following average over $\left| \psi
\left( t\right) \right\rangle $:
\begin{equation}
n_{j,\sigma }\left( t\right) =\frac{1}{N}\left\langle \psi \left( t\right)
\right| a_{j\sigma }^{\dagger }a_{j\sigma }\left| \psi \left( t\right)
\right\rangle ,  \label{eq75}
\end{equation}
we first need to evaluate the averages of the product of operators appearing in the rotated basis, as reported in the Appendix.
Once evaluated the fractions $n_{j,\sigma }\left( t\right) $, with $\sigma =\uparrow
,\downarrow $ and $j=L,R$, whose expression is also reported in the Appendix, we can have access to all physical quantities of interest.

In particular, we get the population imbalance between the two wells
\begin{eqnarray}
D_{LR}\left( t\right) &=&\rho _{\uparrow }\left( t\right) +\rho _{\downarrow
}\left( t\right) =\left( n_{L,\uparrow }\left( t\right) +n_{L,\downarrow
}\left( t\right) \right) -\left( n_{R,\uparrow }\left( t\right)
+n_{R,\downarrow }\left( t\right) \right) =\frac{1}{4}\frac{1}{\overline{%
\Omega }_{\delta }^{N+1}}\cos \left( 2\overline{J}t\right)  \nonumber \\
&&\cdot \left\{ \left[ \left( \overline{\Omega }_{\delta }-\delta \right)
^{2}+\overline{\Omega }^{2}\right] \left[ \frac{\overline{\Omega }_{\delta }%
}{2}\left( 1+\cos \left( \frac{\overline{\Omega }^{2}}{\overline{\Omega }%
_{\delta }^{2}}\frac{\overline{g}t}{4}\right) \right) +\frac{\delta }{2}%
\left( 1-\cos \left( \frac{\overline{\Omega }^{2}}{\overline{\Omega }%
_{\delta }^{2}}\frac{\overline{g}t}{4}\right) \right) \right] ^{N-1}\right.
\label{eq81} \\
&&\left. +\left[ \left( \overline{\Omega }_{\delta }+\delta \right) ^{2}+%
\overline{\Omega }^{2}\right] \left[ \frac{\overline{\Omega }_{\delta }}{2}%
\left( 1+\cos \left( \frac{\overline{\Omega }^{2}}{\overline{\Omega }%
_{\delta }^{2}}\frac{\overline{g}t}{4}\right) \right) -\frac{\delta }{2}%
\left( 1-\cos \left( \frac{\overline{\Omega }^{2}}{\overline{\Omega }%
_{\delta }^{2}}\frac{\overline{g}t}{4}\right) \right) \right] ^{N-1}\right\}
,  \nonumber
\end{eqnarray}
the magnetization
\begin{eqnarray}
M_{LR}\left( t\right) &=&\rho _{\uparrow }\left( t\right) -\rho _{\downarrow
}\left( t\right) =\left( n_{L,\uparrow }\left( t\right) -n_{R,\uparrow
}\left( t\right) \right) -\left( n_{L,\downarrow }\left( t\right)
-n_{R,\downarrow }\left( t\right) \right) =\frac{1}{4}\frac{1}{\overline{%
\Omega }_{\delta }^{N+1}}\cos \left( 2\overline{J}t\right)  \nonumber \\
&&\cdot \left\{ -2\overline{\Omega }^{2}e^{i\overline{\Omega }_{\delta }t}%
\left[ \frac{\overline{\Omega }_{\delta }}{2}\left( \cos \left( \frac{%
\overline{g}\overline{\Omega }_{1}t}{4}\right) +\cos \left( \frac{\overline{g%
}\overline{\Omega }_{2}t}{2}\right) \right) -i\frac{\delta }{2}\left( \sin
\left( \frac{\overline{g}\overline{\Omega }_{1}t}{4}\right) +\sin \left(
\frac{\overline{g}\overline{\Omega }_{2}t}{2}\right) \right) \right]
^{N-1}\right.  \nonumber \\
&&\left. -2\overline{\Omega }^{2}e^{-i\overline{\Omega }_{\delta }t}\left[
\frac{\overline{\Omega }_{\delta }}{2}\left( \cos \left( \frac{\overline{g}%
\overline{\Omega }_{1}t}{4}\right) +\cos \left( \frac{\overline{g}\overline{%
\Omega }_{2}t}{2}\right) \right) +i\frac{\delta }{2}\left( \sin \left( \frac{%
\overline{g}\overline{\Omega }_{1}t}{4}\right) +\sin \left( \frac{\overline{g%
}\overline{\Omega }_{2}t}{2}\right) \right) \right] ^{N-1}\right.
\label{eq82} \\
&&\left. +\left[ \overline{\Omega }^{2}-\left( \overline{\Omega }_{\delta
}-\delta \right) ^{2}\right] \left[ \frac{\overline{\Omega }_{\delta }}{2}%
\left( 1+\cos \left( \frac{\overline{\Omega }^{2}}{\overline{\Omega }%
_{\delta }^{2}}\frac{\overline{g}t}{4}\right) \right) +\frac{\delta }{2}%
\left( 1-\cos \left( \frac{\overline{\Omega }^{2}}{\overline{\Omega }%
_{\delta }^{2}}\frac{\overline{g}t}{4}\right) \right) \right] ^{N-1}\right.
\nonumber \\
&&\left. +\left[ \overline{\Omega }^{2}-\left( \overline{\Omega }_{\delta
}+\delta \right) ^{2}\right] \left[ \frac{\overline{\Omega }_{\delta }}{2}%
\left( 1+\cos \left( \frac{\overline{\Omega }^{2}}{\overline{\Omega }%
_{\delta }^{2}}\frac{\overline{g}t}{4}\right) \right) -\frac{\delta }{2}%
\left( 1-\cos \left( \frac{\overline{\Omega }^{2}}{\overline{\Omega }%
_{\delta }^{2}}\frac{\overline{g}t}{4}\right) \right) \right] ^{N-1}\right\}
,  \nonumber
\end{eqnarray}
and the pseudospin imbalance:
\begin{eqnarray}
D_{\uparrow \downarrow }\left( t\right) &=&\rho _{L}\left( t\right) +\rho
_{R}\left( t\right) =\left( n_{L,\downarrow }\left( t\right)
+n_{R,\downarrow }\left( t\right) \right) -\left( n_{L,\uparrow }\left(
t\right) +n_{R,\uparrow }\left( t\right) \right) =\left\{ \frac{\delta ^{2}}{%
\overline{\Omega }_{\delta }^{2}}+\frac{\overline{\Omega }^{2}}{\overline{%
\Omega }_{\delta }^{N+1}}\frac{e^{i\overline{\Omega }_{\delta }t}}{2}\right.
\nonumber \\
&&\left. \cdot \left[ \frac{\overline{\Omega }_{\delta }}{2}\left( \cos
\left( \frac{\overline{g}\overline{\Omega }_{1}t}{4}\right) +\cos \left(
\frac{\overline{g}\overline{\Omega }_{2}t}{2}\right) \right) -i\frac{\delta
}{2}\left( \sin \left( \frac{\overline{g}\overline{\Omega }_{1}t}{4}\right)
+\sin \left( \frac{\overline{g}\overline{\Omega }_{2}t}{2}\right) \right) %
\right] ^{N-1}+\frac{\overline{\Omega }^{2}}{\overline{\Omega }_{\delta
}^{N+1}}\right.  \label{eq83} \\
&&\left. \cdot \frac{e^{-i\overline{\Omega }_{\delta }t}}{2}\left[ \frac{%
\overline{\Omega }_{\delta }}{2}\left( \cos \left( \frac{\overline{g}%
\overline{\Omega }_{1}t}{4}\right) +\cos \left( \frac{\overline{g}\overline{%
\Omega }_{2}t}{2}\right) \right) +i\frac{\delta }{2}\left( \sin \left( \frac{%
\overline{g}\overline{\Omega }_{1}t}{4}\right) +\sin \left( \frac{\overline{g%
}\overline{\Omega }_{2}t}{2}\right) \right) \right] ^{N-1}\right\},
\nonumber
\end{eqnarray}
where $\overline{\Omega }_{1}=\frac{\overline{\Omega }^{2}-2\delta ^{2}}{%
\overline{\Omega }_{\delta }^{2}}$ and $\overline{\Omega }_{2}=\frac{%
\overline{\Omega }^{2}-\delta ^{2}}{\overline{\Omega }_{\delta }^{2}}$.

One can immediately infer that the temporal modulation of such quantities is much more complicated as it involves more frequencies compared to the characteristic ones $\bar{J}\pm \bar{\Omega}_\delta$ for the non-interacting ones. The atomic and spin currents can be naively obtained by the time derivative of (\ref{eq81})-(\ref{eq82}), respectively.

In Figs. 4, 5 and 6 we show the behaviour of the quantities in Eq.s (\ref{eq81})-(\ref{eq83}) as a function of time for the following choice of parameters: $\overline{\Omega }=0.1 E_{L}$, $\overline{J }=-0.1 E_{L}$, $\overline{g }=0.01$ and $\delta=0.01 E_{L}$, which coincides with the one made in the previous Subsection. Here we switch on the nonlinear interaction, while restricting to the condition $\delta=0.01 E_{L}<<\overline{\Omega }=0.1 E_{L}$ in order to meet the validity range of our analytic calculations. Furthermore we assume a total particle number $N=100$ which is required for the reliability of the two-mode approximation.\\
As expected, quantum collapses and revivals (CR) appear, the whole result of the presence of nonlinearity being a reduction of the oscillation amplitude together with a destruction of periodicity. In particular, in the limit of very small $\delta$ the non-linearity $\bar{g}$ determines the envelope of the revivals as well as the time separation between the adjacent collapse and revival while the separation between neighbouring CRs is proportional to $1/\bar{g}$.
As shown in Fig.s 4-6, with finite $\delta$, the revival occurs at a time scale of the order of tens of ms, which is experimentally accessible. Its observation would be an experimental demonstration of quantum coherence even in the presence of spin-orbit coupling.\\
Another striking feature is the occurrence of a spin Josephson effect,  shown in Fig.5, and which is given by the time-behaviour of the magnetization $M_{LR}$, in full agreement with findings of Ref. \cite{cinesi1}.

\begin{figure}[tbph]
\centering
\includegraphics[scale=0.8]{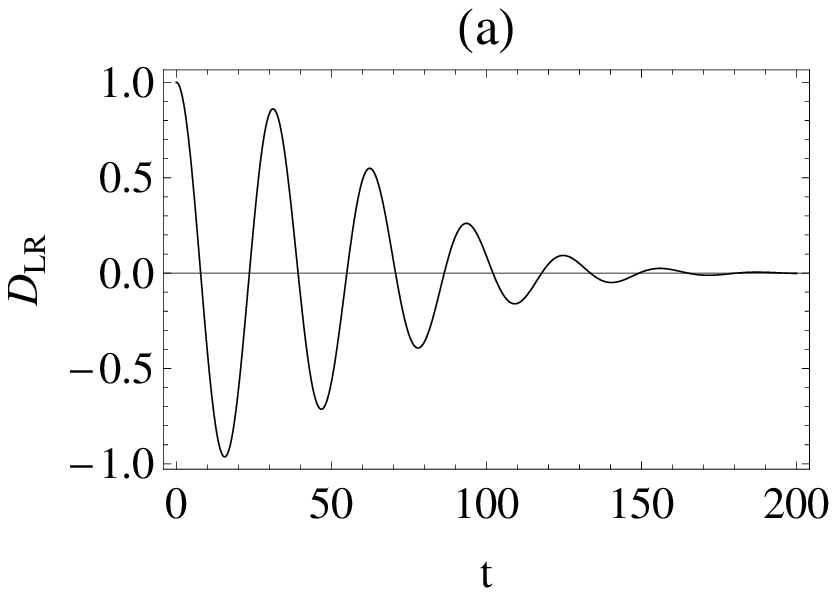}\includegraphics[scale=0.83]{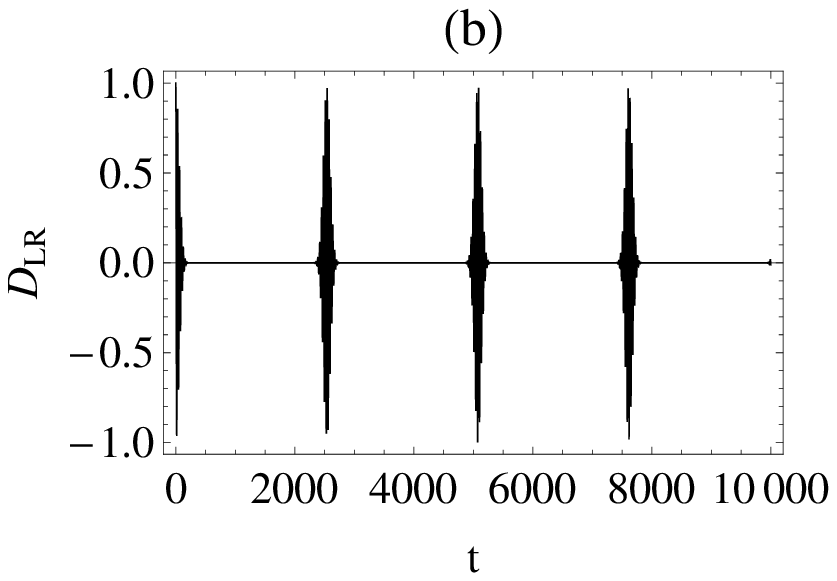}\\
\caption{
Behavior of the population imbalance $D_{LR}$ between the two wells for $N=100$, $\overline{\Omega }=0.1$, $\overline{J }=-0.1$, $\overline{g }=0.01$ and $\delta=0.01$ (units of $E_{L}$). The time is expressed in units of $\frac{\hbar}{E_{L}}$. In the left panel we restrict the time interval to $(0\div 200)$ while the right panel shows collapses and revivals.}\label{fig:fig4}
\end{figure}
\begin{figure}[tbph]
\centering
\includegraphics[scale=0.8]{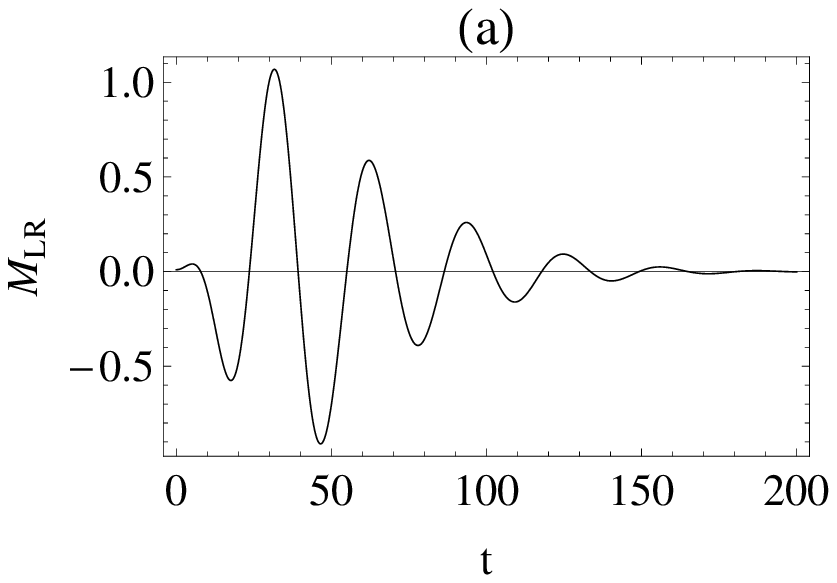}\includegraphics[scale=0.83]{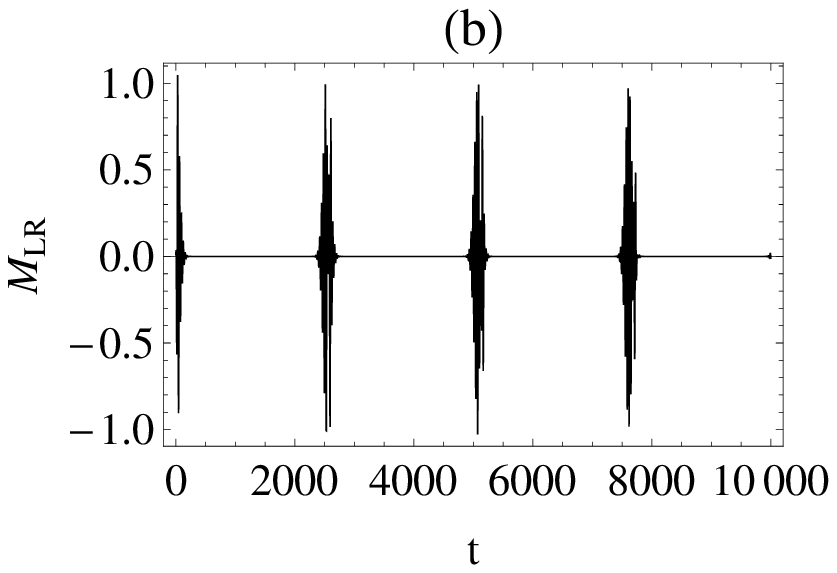}\\
\caption{
Behavior of the magnetization $M_{LR}$ between the two wells for $N=100$, $\overline{\Omega }=0.1$, $\overline{J }=-0.1$, $\overline{g }=0.01$ and $\delta=0.01$ (units of $E_{L}$). The time is expressed in units of $\frac{\hbar}{E_{L}}$. In the left panel we restrict the time interval to $(0\div 200)$ while the right panel shows collapses and revivals.}\label{fig:fig5}
\end{figure}
\begin{figure}[tbph]
\centering
\includegraphics[scale=0.8]{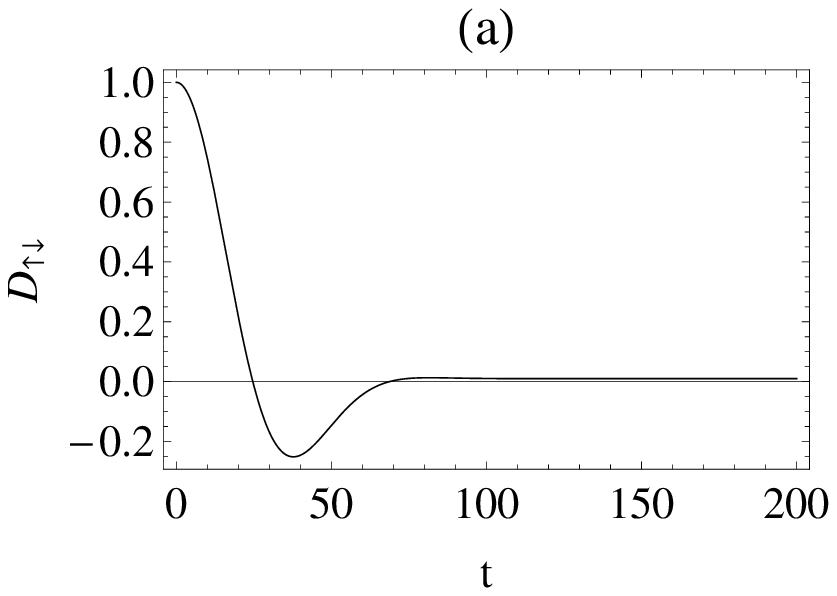}\includegraphics[scale=0.83]{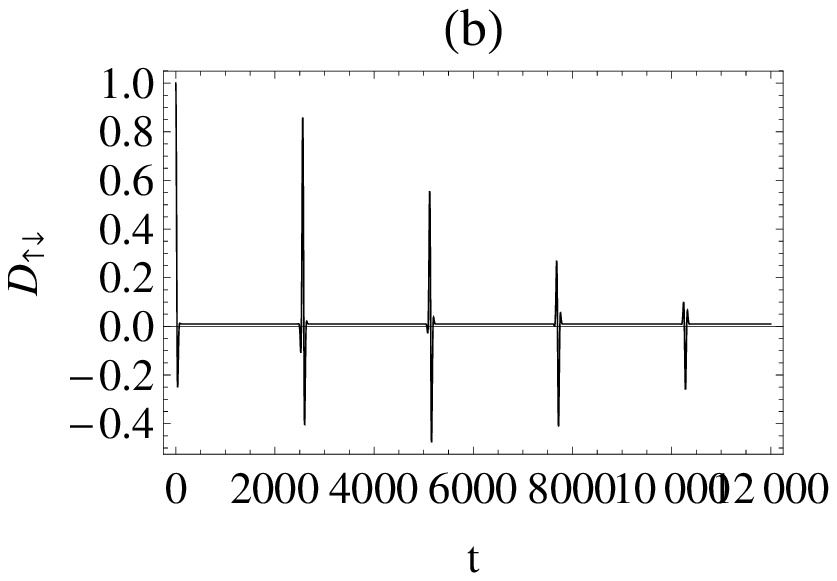}\\
\caption{
Behavior of the pseudospin imbalance $D_{\uparrow \downarrow }$ between the two wells for $N=100$, $\overline{\Omega }=0.1$, $\overline{J }=-0.1$, $\overline{g }=0.01$ and $\delta=0.01$ (units of $E_{L}$). The time is expressed in units of $\frac{\hbar}{E_{L}}$. In the left panel we restrict the time interval to $(0\div 200)$ while the right panel shows collapses and revivals.}\label{fig:fig6}
\end{figure}

From the time derivative of $M_{LR}$  we can numerically evaluate the spin-current $I_s(t)=\frac{d M_{LR}}{dt}$ and define an average spin-current as the integral of $I_s$ over the time interval elapsed between two adjacent collapses and revivals. In Fig. 7 we plot the spin-current as a function of the Zeeman field $\delta$. It shows a linear behavior for small fields $\delta$ and then saturates at higher fields, the linear behavior being characteristic of a non-equilibrium situation.

\begin{figure}[tbph]
\centering
\includegraphics[scale=0.8]{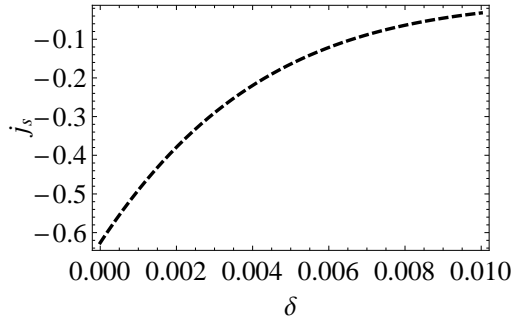}\\
\caption{
Behavior of the spin current $I_s$ flowing between the two wells for $N=100$, $\overline{\Omega }=0.1$, $\overline{J }=-0.1$, $\overline{g }=0.01$ (units of $E_{L}$) as a function of the
Zeeman field. }\label{fig:fig7}
\end{figure}

\section{Conclusions and perspectives}

In this paper we investigated the quantum dynamics of a spin-orbit coupled BEC in a double well potential in a realistic context, by making explicit reference to the experimental setup by NIST group \cite{nist2}. We worked out analytically three different parameters regimes characterized by weak Raman coupling, strong Raman coupling and intermediate coupling respectively. We performed a two-mode approximation and concentrated on the weak interacting regime, which allows a simple analytical study. Indeed our approach doesn't allow one to study the strong nonlinear interaction regime which could show up interesting self-trapping phenomena and is much more amenable to numerical calculations. 
Here the quantum evolution of the number difference of bosons of pseudospin up and down between the two wells is investigated in detail for each parameter regime. Interesting results are found in the intermediate coupling case, both without and with nonlinear interaction; in particular explicit expressions for the time behaviour of the population imbalance between the two wells, the magnetization and the pseudospin imbalance are obtained in correspondence of an initial condition in which all $N$ atoms with pseudospin down in the left well at $t=0$. In the non interacting limit the overall behaviour shows coherent Rabi type oscillations giving rise to an external (population imbalance) and internal (pseudospin imbalance) Josephson effect respectively, while the magnetization exhibits quasiperiodic features. As expected, quantum collapses and revivals appear as a consequence of adding up a weak nonlinear interaction. They occur at a time-scale of the order of tens of ms. Furthermore the time-dependent magnetization $M_{LR}$, which develops in both limits gives rise to a spin Josephson like effect and to a spin current which could be experimentally measured, as shown for instance in Ref. \cite{bloch1},  even if in a different setup (i. e. dynamical control of quantum tunneling in a double wells optical lattice via oscillatory driving fields), and in the more recent paper \cite{bloch2}. Indeed the excellent manipulation of both internal and external degrees of freedom of ultracold atoms could allow one to obtain a net spin current (together with a vanishing atomic current) and to employ it in order to engineer a variety of devices for spintronics \cite{spintronics1}, in analogy with the recently realized atomic counterpart of a spin transistor \cite{spintronics2}.

Last but not least, the ultimate experimental control of the different tunneling processes discussed in this paper could pave the way for the implementation of ultracold atoms analogues of $XXZ$ spin models with tunable couplings \cite{bloch1}.

\section*{Appendix}

The averages of products of the operators appearing in the Hamiltonian (\ref{eq71}) are listed below:
\begin{eqnarray}
\left\langle \psi \left( t\right) \right| c_{1}^{\dagger }c_{1}\left| \psi
\left( t\right) \right\rangle &=&\left\langle \psi \left( t\right) \right|
d_{1}^{\dagger }d_{1}\left| \psi \left( t\right) \right\rangle =\frac{N}{4}%
\left( \frac{\overline{\Omega }_{\delta }-\delta }{\overline{\Omega }%
_{\delta }}\right) ,  \nonumber \\
\left\langle \psi \left( t\right) \right| c_{2}^{\dagger }c_{2}\left| \psi
\left( t\right) \right\rangle &=&\left\langle \psi \left( t\right) \right|
d_{2}^{\dagger }d_{2}\left| \psi \left( t\right) \right\rangle =\frac{N}{4}%
\left( \frac{\overline{\Omega }_{\delta }+\delta }{\overline{\Omega }%
_{\delta }}\right) ,  \nonumber \\
\left\langle \psi \left( t\right) \right| c_{1}^{\dagger }c_{2}\left| \psi
\left( t\right) \right\rangle &=&-e^{i2\omega _{c}t}\frac{N\overline{\Omega }%
}{4\overline{\Omega }_{\delta }^{N}}\left[ \frac{\overline{\Omega }_{\delta }%
}{2}\left( \cos \left( \frac{\overline{g}\overline{\Omega }_{1}t}{4}\right)
+\cos \left( \frac{\overline{g}\overline{\Omega }_{2}t}{2}\right) \right) -i%
\frac{\delta }{2}\left( \sin \left( \frac{\overline{g}\overline{\Omega }_{1}t%
}{4}\right) +\sin \left( \frac{\overline{g}\overline{\Omega }_{2}t}{2}%
\right) \right) \right] ^{N-1},  \nonumber \\
\left\langle \psi \left( t\right) \right| c_{1}^{\dagger }d_{1}\left| \psi
\left( t\right) \right\rangle &=&-e^{i2\overline{J}t}\frac{N\left( \overline{%
\Omega }_{\delta }-\delta \right) }{4\overline{\Omega }_{\delta }^{N}}\left[
\frac{\overline{\Omega }_{\delta }}{2}\left( 1+\cos \left( \frac{\overline{%
\Omega }^{2}}{\overline{\Omega }_{\delta }^{2}}\frac{\overline{g}t}{4}%
\right) \right) +\frac{\delta }{2}\left( 1-\cos \left( \frac{\overline{%
\Omega }^{2}}{\overline{\Omega }_{\delta }^{2}}\frac{\overline{g}t}{4}%
\right) \right) \right] ^{N-1},  \nonumber \\
\left\langle \psi \left( t\right) \right| c_{1}^{\dagger }d_{2}\left| \psi
\left( t\right) \right\rangle &=&e^{i\overline{\Omega }_{\delta }t}\frac{N%
\overline{\Omega }}{4\overline{\Omega }_{\delta }^{N}}\left[ \frac{\overline{%
\Omega }_{\delta }}{2}\left( \cos \left( \frac{\overline{g}\overline{\Omega }%
_{1}t}{4}\right) +\cos \left( \frac{\overline{g}\overline{\Omega }_{2}t}{2}%
\right) \right) -i\frac{\delta }{2}\left( \sin \left( \frac{\overline{g}%
\overline{\Omega }_{1}t}{4}\right) +\sin \left( \frac{\overline{g}\overline{%
\Omega }_{2}t}{2}\right) \right) \right] ^{N-1},  \nonumber \\
\left\langle \psi \left( t\right) \right| c_{2}^{\dagger }d_{1}\left| \psi
\left( t\right) \right\rangle &=&e^{-i\overline{\Omega }_{\delta }t}\frac{N%
\overline{\Omega }}{4\overline{\Omega }_{\delta }^{N}}\left[ \frac{\overline{%
\Omega }_{\delta }}{2}\left( \cos \left( \frac{\overline{g}\overline{\Omega }%
_{1}t}{4}\right) +\cos \left( \frac{\overline{g}\overline{\Omega }_{2}t}{2}%
\right) \right) +i\frac{\delta }{2}\left( \sin \left( \frac{\overline{g}%
\overline{\Omega }_{1}t}{4}\right) +\sin \left( \frac{\overline{g}\overline{%
\Omega }_{2}t}{2}\right) \right) \right] ^{N-1},  \nonumber \\
\left\langle \psi \left( t\right) \right| c_{2}^{\dagger }d_{2}\left| \psi
\left( t\right) \right\rangle &=&-e^{-i2\overline{J}t}\frac{N\left(
\overline{\Omega }_{\delta }+\delta \right) }{4\overline{\Omega }_{\delta
}^{N}}\left[ \frac{\overline{\Omega }_{\delta }}{2}\left( 1+\cos \left(
\frac{\overline{\Omega }^{2}}{\overline{\Omega }_{\delta }^{2}}\frac{%
\overline{g}t}{4}\right) \right) -\frac{\delta }{2}\left( 1-\cos \left(
\frac{\overline{\Omega }^{2}}{\overline{\Omega }_{\delta }^{2}}\frac{%
\overline{g}t}{4}\right) \right) \right] ^{N-1},  \nonumber \\
\left\langle \psi \left( t\right) \right| d_{1}^{\dagger }d_{2}\left| \psi
\left( t\right) \right\rangle &=&-e^{i2\omega _{d}t}\frac{N\overline{\Omega }%
}{4\overline{\Omega }_{\delta }^{N}}\left[ \frac{\overline{\Omega }_{\delta }%
}{2}\left( \cos \left( \frac{\overline{g}\overline{\Omega }_{1}t}{4}\right)
+\cos \left( \frac{\overline{g}\overline{\Omega }_{2}t}{2}\right) \right) -i%
\frac{\delta }{2}\left( \sin \left( \frac{\overline{g}\overline{\Omega }_{1}t%
}{4}\right) +\sin \left( \frac{\overline{g}\overline{\Omega }_{2}t}{2}%
\right) \right) \right] ^{N-1},  \label{eq76}
\end{eqnarray}
where $\overline{\Omega }_{1}=\frac{\overline{\Omega }^{2}-2\delta ^{2}}{%
\overline{\Omega }_{\delta }^{2}}$ and $\overline{\Omega }_{2}=\frac{%
\overline{\Omega }^{2}-\delta ^{2}}{\overline{\Omega }_{\delta }^{2}}$.

In this way, the fraction of pseudospin $\sigma $ in the $j$ well, being $\sigma =\uparrow
,\downarrow $ and $j=L,R$, can be obtained after lengthy but straightforward algebraic calculations:
\begin{eqnarray}
n_{L\downarrow }\left( t\right) &=&\frac{1}{4}\left\{ \frac{1}{2}\left(
\frac{\overline{\Omega }_{\delta }-\delta }{\overline{\Omega }_{\delta }}%
\right) ^{2}+\frac{1}{2}\left( \frac{\overline{\Omega }_{\delta }+\delta }{%
\overline{\Omega }_{\delta }}\right) ^{2}+\frac{\overline{\Omega }^{2}}{%
\overline{\Omega }_{\delta }^{N+1}}\frac{e^{i\overline{\Omega }_{\delta }t}}{%
2}\left( 1+\cos \left( 2\overline{J}t\right) \right) \right.  \nonumber \\
&&\left. \cdot \left[ \frac{\overline{\Omega }_{\delta }}{2}\left( \cos
\left( \frac{\overline{g}\overline{\Omega }_{1}t}{4}\right) +\cos \left(
\frac{\overline{g}\overline{\Omega }_{2}t}{2}\right) \right) -i\frac{\delta
}{2}\left( \sin \left( \frac{\overline{g}\overline{\Omega }_{1}t}{4}\right)
+\sin \left( \frac{\overline{g}\overline{\Omega }_{2}t}{2}\right) \right) %
\right] ^{N-1}\right.  \nonumber \\
&&\left. +\frac{\overline{\Omega }^{2}}{\overline{\Omega }_{\delta }^{N+1}}%
\frac{e^{-i\overline{\Omega }_{\delta }t}}{2}\left( 1+\cos \left( 2\overline{%
J}t\right) \right) \right.  \nonumber \\
&&\left. \cdot \left[ \frac{\overline{\Omega }_{\delta }}{2}\left( \cos
\left( \frac{\overline{g}\overline{\Omega }_{1}t}{4}\right) +\cos \left(
\frac{\overline{g}\overline{\Omega }_{2}t}{2}\right) \right) +i\frac{\delta
}{2}\left( \sin \left( \frac{\overline{g}\overline{\Omega }_{1}t}{4}\right)
+\sin \left( \frac{\overline{g}\overline{\Omega }_{2}t}{2}\right) \right) %
\right] ^{N-1}\right.  \nonumber \\
&&\left. +\frac{\left( \overline{\Omega }_{\delta }-\delta \right) ^{2}}{%
\overline{\Omega }_{\delta }^{N+1}}\frac{\cos \left( 2\overline{J}t\right) }{%
2}\left[ \frac{\overline{\Omega }_{\delta }}{2}\left( 1+\cos \left( \frac{%
\overline{\Omega }^{2}}{\overline{\Omega }_{\delta }^{2}}\frac{\overline{g}t%
}{4}\right) \right) +\frac{\delta }{2}\left( 1-\cos \left( \frac{\overline{%
\Omega }^{2}}{\overline{\Omega }_{\delta }^{2}}\frac{\overline{g}t}{4}%
\right) \right) \right] ^{N-1}\right.  \nonumber \\
&&\left. +\frac{\left( \overline{\Omega }_{\delta }+\delta \right) ^{2}}{%
\overline{\Omega }_{\delta }^{N+1}}\frac{\cos \left( 2\overline{J}t\right) }{%
2}\left[ \frac{\overline{\Omega }_{\delta }}{2}\left( 1+\cos \left( \frac{%
\overline{\Omega }^{2}}{\overline{\Omega }_{\delta }^{2}}\frac{\overline{g}t%
}{4}\right) \right) -\frac{\delta }{2}\left( 1-\cos \left( \frac{\overline{%
\Omega }^{2}}{\overline{\Omega }_{\delta }^{2}}\frac{\overline{g}t}{4}%
\right) \right) \right] ^{N-1}\right\} ,  \label{eq77}
\end{eqnarray}

\begin{eqnarray}
n_{L\uparrow }\left( t\right) &=&\frac{1}{4}\left\{ \left( \frac{\overline{%
\Omega }}{\overline{\Omega }_{\delta }}\right) ^{2}-\frac{\overline{\Omega }%
^{2}}{\overline{\Omega }_{\delta }^{N+1}}\frac{e^{i\overline{\Omega }%
_{\delta }t}}{2}\left( 1+\cos \left( 2\overline{J}t\right) \right) \right.
\nonumber \\
&&\left. \cdot \left[ \frac{\overline{\Omega }_{\delta }}{2}\left( \cos
\left( \frac{\overline{g}\overline{\Omega }_{1}t}{4}\right) +\cos \left(
\frac{\overline{g}\overline{\Omega }_{2}t}{2}\right) \right) -i\frac{\delta
}{2}\left( \sin \left( \frac{\overline{g}\overline{\Omega }_{1}t}{4}\right)
+\sin \left( \frac{\overline{g}\overline{\Omega }_{2}t}{2}\right) \right) %
\right] ^{N-1}\right.  \nonumber \\
&&\left. -\frac{\overline{\Omega }^{2}}{\overline{\Omega }_{\delta }^{N+1}}%
\frac{e^{-i\overline{\Omega }_{\delta }t}}{2}\left( 1+\cos \left( 2\overline{%
J}t\right) \right) \right.  \nonumber \\
&&\left. \cdot \left[ \frac{\overline{\Omega }_{\delta }}{2}\left( \cos
\left( \frac{\overline{g}\overline{\Omega }_{1}t}{4}\right) +\cos \left(
\frac{\overline{g}\overline{\Omega }_{2}t}{2}\right) \right) +i\frac{\delta
}{2}\left( \sin \left( \frac{\overline{g}\overline{\Omega }_{1}t}{4}\right)
+\sin \left( \frac{\overline{g}\overline{\Omega }_{2}t}{2}\right) \right) %
\right] ^{N-1}\right.  \nonumber \\
&&\left. +\frac{\overline{\Omega }^{2}}{\overline{\Omega }_{\delta }^{N+1}}%
\frac{\cos \left( 2\overline{J}t\right) }{2}\left[ \frac{\overline{\Omega }%
_{\delta }}{2}\left( 1+\cos \left( \frac{\overline{\Omega }^{2}}{\overline{%
\Omega }_{\delta }^{2}}\frac{\overline{g}t}{4}\right) \right) +\frac{\delta
}{2}\left( 1-\cos \left( \frac{\overline{\Omega }^{2}}{\overline{\Omega }%
_{\delta }^{2}}\frac{\overline{g}t}{4}\right) \right) \right] ^{N-1}\right.
\nonumber \\
&&\left. +\frac{\overline{\Omega }^{2}}{\overline{\Omega }_{\delta }^{N+1}}%
\frac{\cos \left( 2\overline{J}t\right) }{2}\left[ \frac{\overline{\Omega }%
_{\delta }}{2}\left( 1+\cos \left( \frac{\overline{\Omega }^{2}}{\overline{%
\Omega }_{\delta }^{2}}\frac{\overline{g}t}{4}\right) \right) -\frac{\delta
}{2}\left( 1-\cos \left( \frac{\overline{\Omega }^{2}}{\overline{\Omega }%
_{\delta }^{2}}\frac{\overline{g}t}{4}\right) \right) \right] ^{N-1}\right\}
,  \label{eq78}
\end{eqnarray}
\begin{eqnarray}
n_{R\downarrow }\left( t\right) &=&\frac{1}{4}\left\{ \frac{1}{2}\left(
\frac{\overline{\Omega }_{\delta }-\delta }{\overline{\Omega }_{\delta }}%
\right) ^{2}+\frac{1}{2}\left( \frac{\overline{\Omega }_{\delta }+\delta }{%
\overline{\Omega }_{\delta }}\right) ^{2}+\frac{\overline{\Omega }^{2}}{%
\overline{\Omega }_{\delta }^{N+1}}\frac{e^{i\overline{\Omega }_{\delta }t}}{%
2}\left( 1-\cos \left( 2\overline{J}t\right) \right) \right.  \nonumber \\
&&\left. \cdot \left[ \frac{\overline{\Omega }_{\delta }}{2}\left( \cos
\left( \frac{\overline{g}\overline{\Omega }_{1}t}{4}\right) +\cos \left(
\frac{\overline{g}\overline{\Omega }_{2}t}{2}\right) \right) -i\frac{\delta
}{2}\left( \sin \left( \frac{\overline{g}\overline{\Omega }_{1}t}{4}\right)
+\sin \left( \frac{\overline{g}\overline{\Omega }_{2}t}{2}\right) \right) %
\right] ^{N-1}\right.  \nonumber \\
&&\left. +\frac{\overline{\Omega }^{2}}{\overline{\Omega }_{\delta }^{N+1}}%
\frac{e^{-i\overline{\Omega }_{\delta }t}}{2}\left( 1-\cos \left( 2\overline{%
J}t\right) \right) \right.  \nonumber \\
&&\left. \cdot \left[ \frac{\overline{\Omega }_{\delta }}{2}\left( \cos
\left( \frac{\overline{g}\overline{\Omega }_{1}t}{4}\right) +\cos \left(
\frac{\overline{g}\overline{\Omega }_{2}t}{2}\right) \right) +i\frac{\delta
}{2}\left( \sin \left( \frac{\overline{g}\overline{\Omega }_{1}t}{4}\right)
+\sin \left( \frac{\overline{g}\overline{\Omega }_{2}t}{2}\right) \right) %
\right] ^{N-1}\right.  \nonumber \\
&&\left. -\frac{\left( \overline{\Omega }_{\delta }-\delta \right) ^{2}}{%
\overline{\Omega }_{\delta }^{N+1}}\frac{\cos \left( 2\overline{J}t\right) }{%
2}\left[ \frac{\overline{\Omega }_{\delta }}{2}\left( 1+\cos \left( \frac{%
\overline{\Omega }^{2}}{\overline{\Omega }_{\delta }^{2}}\frac{\overline{g}t%
}{4}\right) \right) +\frac{\delta }{2}\left( 1-\cos \left( \frac{\overline{%
\Omega }^{2}}{\overline{\Omega }_{\delta }^{2}}\frac{\overline{g}t}{4}%
\right) \right) \right] ^{N-1}\right.  \nonumber \\
&&\left. -\frac{\left( \overline{\Omega }_{\delta }+\delta \right) ^{2}}{%
\overline{\Omega }_{\delta }^{N+1}}\frac{\cos \left( 2\overline{J}t\right) }{%
2}\left[ \frac{\overline{\Omega }_{\delta }}{2}\left( 1+\cos \left( \frac{%
\overline{\Omega }^{2}}{\overline{\Omega }_{\delta }^{2}}\frac{\overline{g}t%
}{4}\right) \right) -\frac{\delta }{2}\left( 1-\cos \left( \frac{\overline{%
\Omega }^{2}}{\overline{\Omega }_{\delta }^{2}}\frac{\overline{g}t}{4}%
\right) \right) \right] ^{N-1}\right\} ,  \label{eq79}
\end{eqnarray}

\begin{eqnarray}
n_{R\uparrow }\left( t\right) &=&\frac{1}{4}\left\{ \left( \frac{\overline{%
\Omega }}{\overline{\Omega }_{\delta }}\right) ^{2}-\frac{\overline{\Omega }%
^{2}}{\overline{\Omega }_{\delta }^{N+1}}\frac{e^{i\overline{\Omega }%
_{\delta }t}}{2}\left( 1-\cos \left( 2\overline{J}t\right) \right) \right.
\nonumber \\
&&\left. \cdot \left[ \frac{\overline{\Omega }_{\delta }}{2}\left( \cos
\left( \frac{\overline{g}\overline{\Omega }_{1}t}{4}\right) +\cos \left(
\frac{\overline{g}\overline{\Omega }_{2}t}{2}\right) \right) -i\frac{\delta
}{2}\left( \sin \left( \frac{\overline{g}\overline{\Omega }_{1}t}{4}\right)
+\sin \left( \frac{\overline{g}\overline{\Omega }_{2}t}{2}\right) \right) %
\right] ^{N-1}\right.  \nonumber \\
&&\left. -\frac{\overline{\Omega }^{2}}{\overline{\Omega }_{\delta }^{N+1}}%
\frac{e^{-i\overline{\Omega }_{\delta }t}}{2}\left( 1-\cos \left( 2\overline{%
J}t\right) \right) \right.  \nonumber \\
&&\left. \cdot \left[ \frac{\overline{\Omega }_{\delta }}{2}\left( \cos
\left( \frac{\overline{g}\overline{\Omega }_{1}t}{4}\right) +\cos \left(
\frac{\overline{g}\overline{\Omega }_{2}t}{2}\right) \right) +i\frac{\delta
}{2}\left( \sin \left( \frac{\overline{g}\overline{\Omega }_{1}t}{4}\right)
+\sin \left( \frac{\overline{g}\overline{\Omega }_{2}t}{2}\right) \right) %
\right] ^{N-1}\right.  \nonumber \\
&&\left. -\frac{\overline{\Omega }^{2}}{\overline{\Omega }_{\delta }^{N+1}}%
\frac{\cos \left( 2\overline{J}t\right) }{2}\left[ \frac{\overline{\Omega }%
_{\delta }}{2}\left( 1+\cos \left( \frac{\overline{\Omega }^{2}}{\overline{%
\Omega }_{\delta }^{2}}\frac{\overline{g}t}{4}\right) \right) +\frac{\delta
}{2}\left( 1-\cos \left( \frac{\overline{\Omega }^{2}}{\overline{\Omega }%
_{\delta }^{2}}\frac{\overline{g}t}{4}\right) \right) \right] ^{N-1}\right.
\nonumber \\
&&\left. -\frac{\overline{\Omega }^{2}}{\overline{\Omega }_{\delta }^{N+1}}%
\frac{\cos \left( 2\overline{J}t\right) }{2}\left[ \frac{\overline{\Omega }%
_{\delta }}{2}\left( 1+\cos \left( \frac{\overline{\Omega }^{2}}{\overline{%
\Omega }_{\delta }^{2}}\frac{\overline{g}t}{4}\right) \right) -\frac{\delta
}{2}\left( 1-\cos \left( \frac{\overline{\Omega }^{2}}{\overline{\Omega }%
_{\delta }^{2}}\frac{\overline{g}t}{4}\right) \right) \right] ^{N-1}\right\}
.  \label{eq80}
\end{eqnarray}


\newpage


\begin{thebibliography}{99}
\bibitem{top1}  X. L. Qi, S. C. Zhang, \textit{Physics Today }\textbf{63 }%
(2010) 33; M. Z. Hasan, C. L. Kane, \textit{Rev. Mod. Phys. }\textbf{82 }%
(2010) 3045.

\bibitem{top2}  C. L. Kane, E. J. Mele, \textit{Phys. Rev. Lett. }\textbf{95
}(2005) 146802; B. A. Bernevig, T. L. Hughes, S. C. Zhang, \textit{Science }%
\textbf{314 }(2006) 1757.

\bibitem{top3}  J. D. Sau, R. M. Lutchyn, S. Tewari, S. Das Sarma, \textit{%
Phys. Rev. Lett. }\textbf{104 }(2010) 040502.

\bibitem{cold1}  I. Bloch, J. Dalibard, W. Zwerger, \textit{Rev. Mod. Phys. }%
\textbf{80 }(2008) 885.

\bibitem{cold2}  F. Gerbier, J. Dalibard, \textit{Nature }\textbf{12 }(2010)
033007; J. Dalibard, F. Gerbier, G. Juzeliunas, P. Ohberg, \textit{Rev. Mod.
Phys. }\textbf{83 } (2011) 1523; N. Goldman, G. Juzeliunas, P. Ohberg, I. B.
Spielman, arXiv:1308.6533v1.

\bibitem{nist1}  Y. J. Lin, R. L. Compton, A. R. Perry, W. D. Phillips, J.
V. Porto, I. B. Spielman, \textit{Phys. Rev. Lett. }\textbf{102 } (2009)
130401; Y. J. Lin, R. L. Compton, K. Jimenez-Garcia, J. V. Porto, I. B.
Spielman, \textit{Nature }\textbf{462 } (2009) 628; Y. J. Lin, R. L.
Compton, K. Jimenez-Garcia, W. D. Phillips, J. V. Porto, I. B. Spielman,
\textit{Nat. Phys. }\textbf{7 } (2011) 531.

\bibitem{nist2}  Y. J. Lin, K. Jimenez-Garcia, I. B. Spielman, \textit{%
Nature }\textbf{471 } (2011) 83; V. Galitski, I. B. Spielman, \textit{Nature
}\textbf{494 } (2013) 49.

\bibitem{ra1}  Y. A. Bychkov, E. I. Rashba, \textit{J. Phys. C }\textbf{17 }
(1984) 6039.

\bibitem{ra2}  G. Dresselhaus, \textit{Phys. Rev. }\textbf{100 } (1955) 580.

\bibitem{campbell}  D. L. Campbell, G. Juzeliunas, I. B. Spielman, \textit{%
Phys. Rev. A }\textbf{84 } (2011) 025602.

\bibitem{th1}  C. Wang, C. Gao, C. M. Jian, H. Zhai, \textit{Phys. Rev.
Lett. }\textbf{105 } (2010) 160403; T. L. Ho, S. Zhang, \textit{Phys. Rev.
Lett. }\textbf{107 } (2011) 150403; C. M. Jian, H. Zhai, \textit{Phys. Rev.
B }\textbf{84 } (2011) 060508.

\bibitem{th1bis}  Y. Li, L. P. Pitaevski, S. Stringari, \textit{Phys. Rev.
Lett. }\textbf{108 } (2012) 225301.

\bibitem{th2}  X. Q. Xu, J. H. Han, \textit{Phys. Rev. Lett. }\textbf{107 }
(2011) 200401; X. F. Zhou, J. Zhou, C. J. Wu, \textit{Phys. Rev. A }\textbf{%
84 } (2011) 063624; J. Radic, T. A. Sedrakyan, I. B. Spielman, V. Galitski,
\textit{Phys. Rev. A }\textbf{84 } (2011) 063604.

\bibitem{th3}  E. van der Bijl, R. A. Duine, \textit{Phys. Rev. Lett. }%
\textbf{107 } (2011) 195302.

\bibitem{th3bis}  Y. Li, G. I. Martone, S. Stringari, \textit{Europhys.
Lett. }\textbf{99 } (2012) 56008.

\bibitem{th4}  T. Grass, K. Saha, K. Sengupta, M. Lewenstein, \textit{Phys.
Rev. A }\textbf{84 } (2011) 053632.

\bibitem{th4bis}  Y. Li, G. I. Martone, L. P. Pitaevski, S. Stringari,
\textit{Phys. Rev. Lett. }\textbf{110 } (2013) 235302.

\bibitem{th5}  Y. Zhang, L. Mao, C. Zhang, \textit{Phys. Rev. Lett. }\textbf{%
108 } (2012) 035302.

\bibitem{th6}  D. W. Zhang, Z. Y. Xue, H. Yan, Z. D. Wang, S. L. Zhu,
\textit{Phys. Rev. A }\textbf{85 } (2012) 013628.

\bibitem{th7}  H. Zhai, \textit{Int. J. Mod. Phys. B }\textbf{26 } (2012)
1230001.

\bibitem{leggett1}  A. J. Leggett, \textit{Rev. Mod. Phys. }\textbf{73 }%
(2001) 307.

\bibitem{junction1}  B. D. Josephson, \textit{Phys. Lett. }\textbf{1 }(1962)
251; B. D. Josephson, \textit{Rev. Mod. Phys. }\textbf{36 }(1964) 216.

\bibitem{jos1}  J. Javanainen, \textit{Phys. Rev. Lett. }\textbf{57 }(1986)
3164; I. Zapata, F. Sols, A. J. Leggett, \textit{Phys. Rev. A }\textbf{57 }%
(1998) 1050.

\bibitem{milburn}  G. J. Milburn, J. Corney, E. M. Wright, D. F. Walls,
\textit{Phys. Rev. A }\textbf{55 }(1997) 4318.

\bibitem{ananikian1}  D. Ananikian, T. Bergeman, \textit{Phys. Rev. A }%
\textbf{74 }(2006) 039905.

\bibitem{smerzi1}  A. Smerzi, S. Fantoni, S. Giovanazzi, S. R. Shenoy,
\textit{Phys. Rev. Lett. }\textbf{79 }(1997) 4950; S. Raghavan, A. Smerzi,
S. Fantoni, S. R. Shenoy, \textit{Phys. Rev. A }\textbf{59 }(1999) 620.

\bibitem{smerzi2}  S. Giovanazzi, A. Smerzi, S. Fantoni, \textit{Phys. Rev.
Lett. }\textbf{84 }(2000) 4521.

\bibitem{exp1}  F. S. Cataliotti, S. Burger, C. Fort, P. Maddaloni, F.
Minardi, A. Trombettoni, M. Inguscio, \textit{Science }\textbf{293 }(2001)
843.

\bibitem{exp2}  M. Albiez, R. Gati, J. Folling, S. Hunsmann, M. Cristiani,
M. K. Oberthaler, \textit{Phys. Rev. Lett. }\textbf{95 }(2005) 010402.

\bibitem{exp3}  S. Levy, E. Lahoud, I. Shomroni, J. Steinhauer, \textit{%
Nature }\textbf{449 }(2007) 579.

\bibitem{dimer1}  G. Kalosakas, A. R. Bishop, \textit{Phys. Rev. A }\textbf{%
65 }(2002) 043616; G. Kalosakas, A. R. Bishop, V. M. Kenkre, \textit{Phys.
Rev. A }\textbf{68 }(2003) 023602.

\bibitem{gati1}  R. Gati, M. K. Oberthaler, \textit{J. Phys. B: At. Mol.
Opt. }\textbf{40 }(2007) R61.

\bibitem{ferrini1}  G. Ferrini, A. Minguzzi, F. W. J. Hekking, \textit{Phys.
Rev. A }\textbf{78 }(2008) 023606.

\bibitem{lipkin1}  H. J. Lipkin, N. Meshkov, A. J. Glick, \textit{Nucl.
Phys. }\textbf{62 }(1965) 188; N. Meshkov, A. J. Glick, H. J. Lipkin,
\textit{Nucl. Phys. }\textbf{62 }(1965) 199; A. J. Glick, H. J. Lipkin, N.
Meshkov, \textit{Nucl. Phys. }\textbf{62 }(1965) 211.

\bibitem{lipkin2}  S. Dusuel, J. Vidal, \textit{Phys. Rev. B }\textbf{71 }%
(2005) 224420; P. Ribeiro, J. Vidal, R. Mosseri, \textit{Phys. Rev. Lett. }%
\textbf{99 }(2007) 050402; R. Orus, S. Dusuel, J. Vidal, \textit{Phys. Rev.
Lett. }\textbf{101 }(2008) 025701.

\bibitem{mix0}  S. Ashhab, C. Lobo, \textit{Phys. Rev. A }\textbf{66 }(2002)
013609; H. Pu, W. Zhang, P. Meystre, \textit{Phys. Rev. Lett. }\textbf{89 }%
(2002) 090401; K. Molmer, \textit{Phys. Rev. Lett. }\textbf{90 }(2003)
110403.

\bibitem{hp3}  H. T. Ng, P. T. Leung, \textit{Phys. Rev. A }\textbf{71 }%
(2005) 013601.

\bibitem{mix1}  G. Mazzarella, M. Moratti, L. Salasnich, M. Salerno, F.
Toigo, \textit{J. Phys. B: At. Mol. Opt. }\textbf{42 }(2009) 125301.

\bibitem{mix2}  X. Q. Xu, L. H. Lu, Y. Q. Li, \textit{Phys. Rev. A }\textbf{%
78 }(2008) 043609.

\bibitem{mix3}  I. I. Satija, R. Balakrishnan, P. Naudus, J. Heward, M.
Edwards, C. W. Clark, \textit{Phys. Rev. A }\textbf{79 }(2009) 033616.

\bibitem{mix4}  B. Julia-Diaz, M. Guilleumas, M. Lewenstein, A. Polls, A.
Sanpera, \textit{Phys. Rev. A }\textbf{80 }(2009) 023616.

\bibitem{mix4bis}  M. Guilleumas, B. Julia-Diaz, M. Mele-Messeguer, A.
Polls, \textit{Las. Phys. }\textbf{20 }(2010) 1163.

\bibitem{mix5}  M. Mele-Messeguer, B. Julia-Diaz, M. Guilleumas, A. Polls,
A. Sanpera, \textit{New J. Phys. }\textbf{13 }(2011) 033012.

\bibitem{noi1}  A. Naddeo, R. Citro, \textit{J. Phys. B: At. Mol. Opt. }%
\textbf{43 }(2010) 135302.

\bibitem{noi2}  R. Citro, A. Naddeo, E. Orignac, \textit{J. Phys. B: At.
Mol. Opt. }\textbf{44 }(2011) 115306.

\bibitem{mix6}  B. Sun, M. S. Pindzola, \textit{Phys. Rev. A }\textbf{80 }%
(2009) 033616.

\bibitem{cinesi1}  D. W. Zhang, L. B. Fu, Z. D. Wang, S. L. Zhu, \textit{%
Phys. Rev. A }\textbf{85 }(2012) 043609.

\bibitem{MDS1}  M. A. Garcia-March, G. Mazzarella, L. Dell'Anna, B. Julia-Diaz, L. Salasnich, A. Polls, arXiv:1401.7693v1.

\bibitem{hp1}  T. Holstein, H. Primakoff, \textit{Phys. Rev. }\textbf{58 }%
(1949) 1098.

\bibitem{hp2}  M. P. Strzys, J. R. Anglin, \textit{Phys. Rev. A }\textbf{81 }%
(2010) 043616.

\bibitem{rw1}  A. P. Alodjanc, S. M. Arakeljan, A. S. Chirkin, \textit{JETP }%
\textbf{81 }(1995) 34.

\bibitem{rw2}  L. M. Kuang, Z. W. Ouyang, \textit{Phys. Rev. A }\textbf{80 }%
(2009) 033616.


\bibitem{bloch1}  Y. A. Chen, S. Nascimbene, M. Aidelsburger, M. Atala, S. Trotzky, I. Bloch, \textit{Phys. Rev. Lett. }\textbf{107 }%
(2011) 210405.

\bibitem{bloch2} M. Atala, M. Aidelsburger, M. Lohse, J. T. Barreiro, B. Paredes, I. Bloch, arXiv:1402.0819v1.

\bibitem{spintronics1}  I. Zutic, J. Fabian, S. Das Sarma, \textit{Rev. Mod. Phys. }\textbf{76 }%
(2004) 323.

\bibitem{spintronics2}  M. C. Beeler, R. A. Williams, K. Jimenez-Garcia, L. J. Le Blanc, A. R. Perry, I. B. Spielman, \textit{Nature }\textbf{498 }%
(2013) 201.
\end{thebibliography}
\end{document}